\documentclass[a4,final]{elsarticle}

\usepackage{amssymb}
\usepackage{amsthm}
\usepackage{amsmath}

\biboptions{sort&compress}

\renewcommand{\vec}[1]{\mathbf{#1}}
\newcommand{\gvec}[1]{\boldsymbol{#1}}
\newcommand{\vor}{\gvec{\omega}}
\newcommand{\vel}{\vec{u}}
\newcommand{\x}{\vec{x}}
\newcommand{\y}{\vec{y}}
\newcommand{\X}{\vec{X}}
\newcommand{\Alpha}{\gvec{\alpha}}
\newcommand{\vxi}{\gvec{\xi}}

\begin{document}
\begin{frontmatter}

\title{Lagrangian and geometric analysis of finite-time Euler
  singularities}

\author[rub]{T. Grafke}
\ead{tg@tp1.rub.de}
\author[rub]{R. Grauer}
\ead{grauer@tp1.rub.de}
\address[rub]{Institut f\"ur Theoretische Physik I, Ruhr-Universit\"at Bochum, Germany}

\begin{abstract}
  We present a numerical method of analyzing possibly singular
  incompressible 3D Euler flows using massively parallel
  high-resolution adaptively refined numerical simulations up to
  $8192^3$ mesh points. Geometrical properties of Lagrangian vortex
  line segments are used in combination with analytical non-blowup
  criteria by Deng et al [Commun. PDE \textbf{31} (2006)] to reliably
  distinguish between singular and near-singular flow evolution. We
  then apply the presented technique to a class of high-symmetry
  initial conditions and present numerical evidence against the
  formation of a finite-time singularity in this case.
\end{abstract}

\begin{keyword}
  Euler equation \sep Existence, uniqueness and regularity theory \sep Vortex line geometry \sep Adaptive mesh refinement


\end{keyword}

\end{frontmatter}

\section{Introduction}

For now more than two centuries, the incompressible Navier-Stokes
equations have withstood the minds of mathematicians and physicists
alike: The derivation of the nature of turbulence from the equations,
as well as the global existence of smooth solutions is not known to
date. The huge mathematical difficulties concerning the latter problem
were recognized by its elevation to the status of ``Millennium Prize
Problem'' by the Clay Mathematics Institute (see the official problem
description by Fefferman \cite{fefferman:2000}, or review articles
e.g. \cite{ladyzhenskaya:2003,doering:2009}). A proof of existence of
global regular solutions to the Navier-Stokes equation is believed to
entail the development of completely new methods for the analysis of
partial differential equations. The absence of mathematical certainty
for the Navier-Stokes equations may seem to leave the physicist in a
somewhat embarrassing position: The equation is known and well tested
in application, but the existence of solutions is unclear in relevant
cases.

Yet, the actual impact of a supposed breakdown of solutions for the
Navier-Stokes equations on physics of fluids is smaller than one might
expect and appears like a mere technical detail on second
thought. Singularities in the Navier-Stokes equation would, if
existent, appear on very small scales. Obviously, continuum mechanics
do not hold on these smallest scales and the breakdown of the model
equation would appear in a regime in which the model does not describe
reality at any rate. Furthermore, the nature of supposed singularities for
the Navier-Stokes equation is proven to be unphysical in nature, as it
requires the existence of infinite momentum. Without external forcing,
from smooth initial conditions and in the presence of friction, the
occurrence of infinite momentum is impossible to justify
physically. The impact of singularities is additionally limited by the
fact that the space-time dimension of the singular region is proven to
be less than or equal to one for the Navier-Stokes equations
\cite{caffarelli-kohn-nirenberg:1982}.

In the inviscid limit, the situation is quite the opposite. The
incompressible Euler equations for ideal fluids,
\begin{equation}
  \label{eq:euler}
  \frac{\partial \vel}{\partial t} + \vel \cdot \nabla \vel + \nabla p = 0\,, \quad \nabla \cdot \vel = 0\,.
\end{equation}
appear to be of little physical significance in most applications,
since friction is the dominating process on small scales. The
ignorance regarding existence of global solutions is even larger for
the inviscid case: The notion of weak solutions, which are well
established for the Navier-Stokes equations since Leray
\cite{leray:1934}, is unknown for the three-dimensional Euler
equations. Nevertheless, the formation of finite-time Euler
singularities significantly concerns our understanding of (viscid)
fluid dynamics. Euler singularities, if existent, would coincide with
the development of large gradients in the velocity field. Since no
friction is limiting the increase in velocity gradients, infinite
momentum is not mandatory for a blowup of the Euler equations. The
inviscid limit is, therefore, not only a mere description of ideal
fluids, but explores the possibility of inherent dynamical processes
beyond friction that limit the transition to smaller and smaller
scales. This has immediate implications on the existence of a cut-off
velocity in high Reynolds-number Navier-Stokes flows, leading to the
slightly exaggerated question, quoting Constantin
\cite{constantin:2007}: ``Do we need Schr\"odinger’s equations to
calculate the flow around a moving car? Or to predict tomorrow's
weather?'' For that reason, the problem of singularities for the Euler
equations is of far greater importance to the physical understanding
of fluids than the analogous problem for the Navier-Stokes equations.

A similar argument is valid for turbulence. Today's phenomenological
description of turbulence (e.g. \cite{she-leveque:1994,frisch:1995}),
which is built on the basis of the celebrated theory by Kolmogorov
\cite{kolmogorov:1941,kolmogorov:1941b,kolmogorov:1962}, contains as a
central point that, in the limit of vanishing viscosity, energy
dissipation has to stay finite. This behavior could be explained by
the formation of finite-time Euler singularities, as implied by
Onsager's conjecture \cite{onsager:1949}. For three-dimensional
incompressible flow, non-conservation of energy might be caused not
only by viscosity but by missing regularity in the velocity
field. Energy dissipation might occur, if the H\"older continuity
exponent is smaller than $1/3$ for the velocity field. This conjecture
was proven in terms of Besov spaces \cite{constantin-e-titi:1994,
  cheskidov-constantin-friedlander-shvydkoy:2008}. As a consequence, a
mathematical description of turbulence might be possible in terms of
weak solutions for the Euler equations, if smooth solutions gain
enough roughness in finite time. Therefore, insight into the formation
of finite-time singularities for the Euler equations could uncover a
mechanism essential for the understanding of viscous turbulence.

The search for finite-time singularities of the Euler equations has
resulted in extensive literature, with many analytical results being
relatively young. Especially the advent of scientific computing has
given research a new direction: Reports of numerical evidence
supporting or denying the existence of finite-time singularities for
the Euler equations are numerous (see e.g. \cite{gibbon:2008} for a
compiled list).

As a now classical result, the blowup criterion of Beale et. al
\cite{beale-kato-majda:1984} (BKM) connects the existence of solutions
for the incompressible Euler equations in three dimensions to the
critical accumulation of vorticity. More recently, geometric analysis
of the flow
\cite{constantin-fefferman-majda:1996,cordoba-fefferman:2001} has
helped increasing insight into the process of vorticity growth. Among
these geometric blowup criteria, theorems developed by Deng et. al
\cite{deng-hou-yu:2005} may be seen as the first to be suitable for
verification by direct numerical simulations. An approach along this
way will be presented in this paper.

Given the results of analytical considerations and the experience
gained from numerical simulation of the Euler equations, certain
scenarios are known to be possibly compatible with the analytic
requirements of a finite-time blowup, namely the global notion of
self-similar collapse to a point and the local process of vorticity
accumulation by vorticity-strain coupling. It has been tried in the
past to construct explicit initial conditions exploiting these
scenarios to obtain numerical evidence for or against a finite-time
singularity, with surprisingly inconsistent results. The major reason
for this ambiguity is the critical dependence on extrapolation, which
renders the identification of singular versus near-singular behavior
next to impossible by numerical means. The hopes are high that the
situation is less vague when considering geometric properties of the
flow, as mentioned above. We will present the application of such
geometric criteria to numerical data to sharpen the distinction
between singular and near-singular flow evolution and identify the
processes connected with this behavior.

This paper is organized as follows: we first review the relevant
geometric blowup-criteria that form the basis of our numeric
method. This includes a motivation of how geometric properties such as
curvature or spreading of Lagrangian vortex line segments are
connected to the accumulation of vorticity. We then describe the
special class of high symmetry initial conditions used for our
numerical experiments, discuss the implications of flow symmetries on
the process of vorticity-strain coupling and introduce different
vorticity profiles for vortex dodecapole initial conditions. Using
this setup, details of the its implementation for our massively
parallel simulations with up to $8192^3$ mesh points are
given. Results are presented concerning the growth of vorticity and
strain, the BKM-criterion and the geometric criteria. These findings
act as numerical evidence against the formation of a finite-time
singularity for this class of initial conditions.  A conclusion and
outlook summarize the paper.

\section{Geometric blowup criteria}
\label{sec:geometric}

Classical criteria for the development of a finite-time Euler
singularity have in common that they focus on global features (such as
certain norms of the velocity or the vorticity fields) or on
point-wise Eulerian features (such as $\Omega(t)$) of the flow. This
comes at the disadvantage of neglecting the structures and physical
mechanisms of the flow evolution. A strategy to overcome such
shortcomings was established by focusing more on geometrical
properties and flow structures, such as vortex tubes or vortex
lines. Starting with the works of Constantin et al.
\cite{constantin:1994,constantin-fefferman-majda:1996} and Cordoba and
Fefferman \cite{cordoba-fefferman:2001}, some of these ``geometric''
criteria (e.g. \cite{gibbon:2002, deng-hou-yu:2005,
  gibbon-holm-kerr-roulstone:2006}) have reached a phase where they
allow direct verification of their assumptions with the help of
numerical simulations. Special focus is placed on the criteria
presented by Deng, Hou and Yu \cite{deng-hou-yu:2005,
  deng-hou-yu:2006}, as the assumptions are in close reach for
numerical simulations. They examine the Lagrangian evolution of vortex
line segments and formulate a combined bound on velocity blowup and
vortex segment collapse.

One of the trivial consequences of the BKM theorem is the fact that no
blowup can occur for the two dimensional Euler equations. Since the
vorticity $\vor(\x,t)$ is bounded by the initial conditions
$\|\vor_0\|_{L^\infty}$ for all times, a critical accumulation is
impossible. This is a direct consequence of the vorticity pointing out
of the plane of motion, therefore having the vortex-stretching term
$\vor \cdot \nabla \vel$ vanish everywhere. This may be interpreted as
a motivation to focus on the behavior of the direction of vorticity,
$\vxi=\vor/|\vor|$ in the three-dimensional case. For 2D, $\vxi$ is a
constant in space and time (neglecting sign). In 3D, the consequences
of the regularity of $\vxi$ on the growth-rate of vorticity and
ultimately of the applicability of BKM can be precisely stated.

For the Euler equations, this was introduced by Constantin et
al. \cite{constantin-fefferman-majda:1996}. They state, roughly, that
for a smoothly directed vorticity in an $O(1)$-region there may be no
blowup in finite time as long as the velocity remains finite in this
region. Even though this criterion takes into account the local
structure of the flow and follows the evolution of vortex lines, the
(global) bound on the velocity makes this theorem hard to apply in
practice. Numerical simulations of the Euler equations give no
evidence for the velocity to be uniformly bounded in time. This
restriction on the velocity field is weakened in a similar criterion
by Cordoba and Fefferman \cite{cordoba-fefferman:2001}. They consider
vortex tubes with some properties concerning their regularity and a
surrounding $O(1)$ region $Q$ of the flow. From this it is possible to
deduce, with the help of a milder assumption on the surrounding
velocity, that the vortex tube cannot reach zero thickness in finite
time. Even though the velocity field is no longer required to be
uniformly bounded in time, the notion of ``regular tube'' of $O(1)$
length is too restricting, compared to the experiences of numerical
simulations.

\subsection{Regularity of vorticity direction along a vortex line}

Vortex lines of the three-dimensional incompressible Euler equations,
defined as integral curves of the vorticity direction field, are
transported with the flow. As a consequence, two points $\x$ and $\y$ on
the same vortex line $c(s)$ stay on the same vortex line for all
times. Furthermore, as a direct implication of the solenoidality of
the vorticity vector field, the vorticity flux through a vortex tube
is the same for each cross-section.

These two arguments may be combined to get a differential notion of
the connection between the vorticity at two different points on the
same vortex line. A simple consequence of the solenoidality of $\vor$
results in
\begin{equation}
  (\vxi \cdot \nabla) |\vor| = - |\vor| (\nabla \cdot \vxi)\;.
  \label{eq:t1_1}
\end{equation}
Since for a vortex line $c(s)$ it holds by definition that $\dot{c}(s)
= \vxi(c(s))$, we have $\vxi \cdot \nabla \equiv \partial/\partial s$,
where $\partial/\partial s$ is the partial derivative in direction of
the vortex line. Thus, integrating eq. (\ref{eq:t1_1}) along the
vortex line yields
\begin{equation}
  \label{eq:vorticityonaline}
  |\vor(\vec{y}(t),t)| = |\vor(\x(t),t)| \, \exp\left(-\int_{\x(t)}^{\vec{y}(t)} \nabla \cdot \vxi \mathrm{d}s \right)\;.
\end{equation}
Paraphrased, this means: The vorticity at two different points on the
same vortex line is connected by the amount of converging or diverging
of neighboring vortex lines along their interconnecting path. The more
violent vortex lines converge around a vortex line, the faster the
vorticity grows along that line.

\begin{figure}[t]
  \centering
  \includegraphics[width=0.45\linewidth]{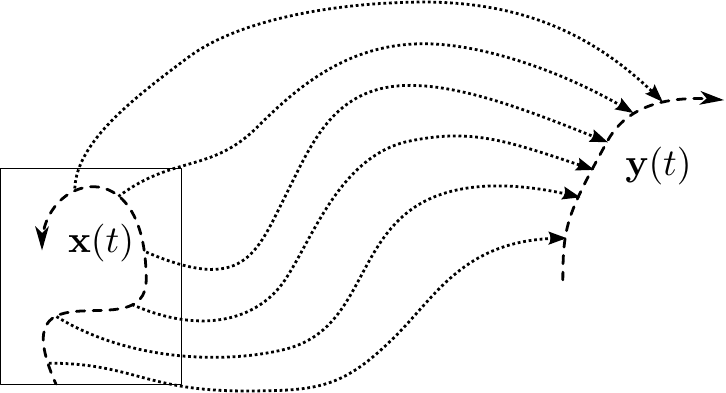}\hspace{0.08\linewidth}
  \includegraphics[width=0.45\linewidth]{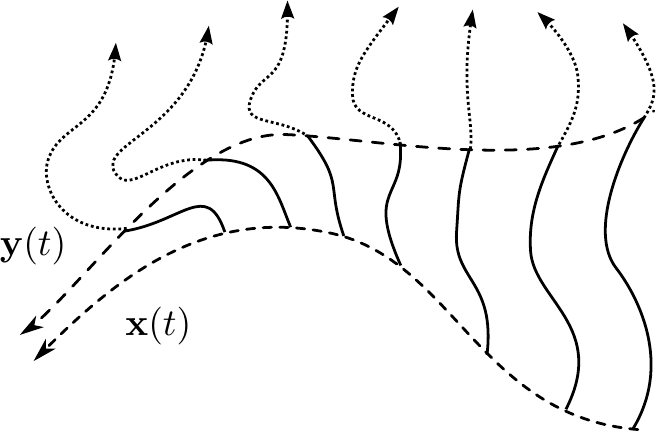}
  \caption{Two ways to apply theorem 1. Left: Choose $\y(t)$ such that
    it is far outside of the critical region of maximum vorticity and
    monitor the behavior of $\int_{\x(t)}^{\y(t)} \left( \nabla \cdot
    \vxi \right) (c(s),t) \,\mathrm{d}s$. Right: For the position
    $\x(t)$ of maximum vorticity, choose $\y(t)$ such that
    $\int_{\x(t)}^{\y(t)} \left( \nabla \cdot \vxi \right) (c(s),t)
    \,\mathrm{d}s = C$. For a point-wise singularity, $\x(t)$ and
    $\y(t)$ must collapse in finite time. \label{fig:T1}}
\end{figure}
This finding was connected with BKM by Deng et
al. \cite{deng-hou-yu:2005} to formulate a geometric blowup
criterion. It is obvious from equation (\ref{eq:vorticityonaline})
that the maximum vorticity $\Omega(t)$ at a given time $t$ can be
estimated by the vorticity on its vortex line, as long as $\nabla
\cdot \vxi$ remains finite. In detail this means:

 \vspace{0.5ex} \textbf{Deng-Hou-Yu theorem 1:} \textit{Let $\x(t)$ be
  a family of points such that for some $c_0>0$ it holds
  $|\omega(\x(t),t)|>c_0\Omega(t)$. Assume that for all $t \in [0,T)$
  there is another point $\y(t)$ on the same vortex line as $\x(t)$,
  such that the direction of vorticity $\vxi(\x,t) =
  \omega(\x,t)/|\omega(\x,t)|$ along the vortex line $c(s)$ between
  $\x(t)$ and $\y(t)$ is well-defined. If we further assume that}
   \begin{equation}
     \left| \;\int_{\x(t)}^{\y(t)} \left( \nabla \cdot \vxi \right) (c(s),t) \,\mathrm{d}s \;\right| \leq C
   \end{equation}
   \textit{for some absolute constant C, and}
   \begin{equation}
     \int_0^T |\omega(\y(t),t)|\,\mathrm{d}t < \infty\;,
   \end{equation}
   \textit{then there will be no blowup up to time $T$.}
 \vspace{0.5ex}

It is immediately clear how this criterion can be applied to numerical
simulations: If the maximum vorticity $\Omega(t)$ exhibits fast growth
in time for which it is hard to decide whether it is a finite-time
blowup compatible with BKM, instead one could monitor the vorticity
outside the critical region, but on the same vortex line. If it
remains well bounded, and $\nabla \cdot \vxi$ along the vortex line
does not scale critically in time, it is safe to deduce a non-critical
growth of $\Omega(t)$. This approach is sketched in Fig. \ref{fig:T1}
(left).

However, due to the freedom of the choice of $\y(t)$ on the critical
vortex line, theorem 1 may be employed in a different way to
distinguish different scenarios for a finite-time singularity, as
depicted in Fig. \ref{fig:T1} (right). Suppose that the maximum
vorticity $\Omega(t)$ grows in a way compatible with BKM. Now, choose
$\x(t)$ to be the position of maximum vorticity and define $\y(t)$ via
\begin{equation}
  \int_{\x(t)}^{\y(t)} \nabla \cdot \vxi \mathrm{d}s = C
\end{equation}
for some constant $C$ independent of the time $t$, where $s$ denotes
the arc-length parameter of the curve from $\x(t)$ to $\y(t)$. In
words, choose $\y(t)$ on the same vortex line as $\x(t)$ such that the
accumulation of tightening of nearby vortex lines is the same for every
instance in time. This provides us with the ability to clearly
distinguish between to separate cases of supposed blowup:
\begin{enumerate}[(i)]
\item For every constant $C$, $\y(t)$ approaches $\x(t)$ in finite
  time to collapse to a single point. This would constitute the
  desired behavior for a point-wise singularity in the origin.
\item If for any constant $C$, $\x(t)$ and $\y(t)$ stay well separated
  in time and do not collapse to a point, the whole vortex-line from
  $\x(t)$ to $\y(t)$ has to blow up in order to maintain critical
  growth in $\x(t)$. This scenario, however unlikely, is not
  ruled out analytically.
\end{enumerate}
The insight provided by the theorem could successfully be used as
evidence excluding a point-wise singularity for the considered initial
conditions as presented below.

\subsection{Vortex line stretching and vorticity accumulation}
\label{ssec:vortexlinestretching}

\begin{figure}[t]
  \centering
  \includegraphics[width=0.3\textwidth]{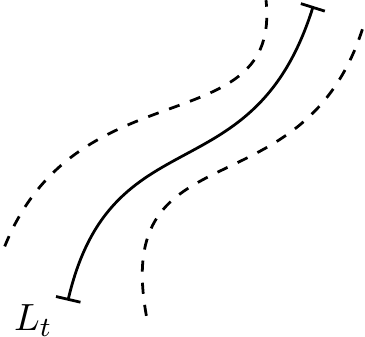}
  \includegraphics[width=0.3\textwidth]{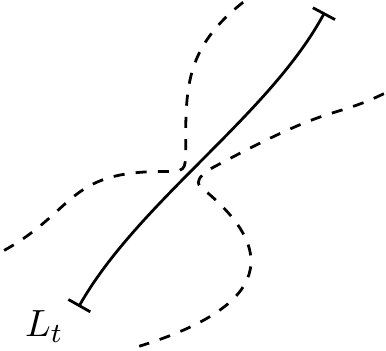}
  \includegraphics[width=0.3\textwidth]{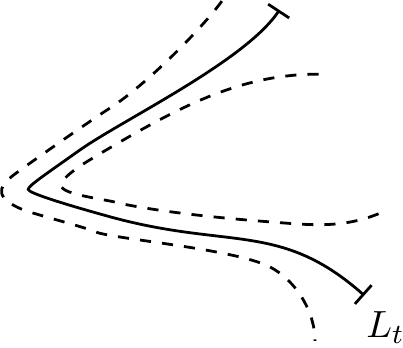}
  \caption[Characterizing vortex line geometry by
    $\lambda(L_t)$]{Characterizing vortex line geometry in terms of
    $\lambda(L_t)$. A slightly curved vortex line with approximately
    parallel neighboring vortex lines (left) exhibits small
    $\lambda(L_t)$. Vortex lines with tightening neighboring vortex
    lines (center) or vortex lines with high curvature, in
    comparison to their length (right) have high
    $\lambda(L_t)$.}
  \label{fig:T2lambda}
\end{figure}
Vortex stretching is recognized as the mechanism responsible for the
accumulation of vorticity. Revisited from a geometric point of view,
vortex lines are transported with the flow, yet twist and turn due to
vortex stretching. Since in the absence of dissipation vortex lines
are unable to reconnect, the topological properties of vortex lines
are fixed. A complex flow will therefore entangle, stretch and twist
vortex lines in a non-trivial way.

The geometric equivalent of the vortex stretching term is the increase
in length for a Lagrangian vortex line. It is possible to quantify
this stretching and establish a sound connection to the vorticity
dynamics of the flow. This in turn can then be used to reformulate
blowup criteria in terms of geometric constraints on Lagrangian vortex
lines. This section is meant to give an overview over this procedure
to understand the implications of the second theorem of
\cite{deng-hou-yu:2005}. Details regarding the statements below are
given therein.

Consider a vortex line segment $L_0$ at time $t=0$ and its Lagrangian
image $L_t = \X(L_0,t)$. Let $\beta$, $s$ be the arc length parameters
of $L_t$ at times $0$ and $t$. Then, a direct implication of the
vorticity transport formula describes the evolution of the absolute
vorticity at a Lagrangian fluid element
\begin{align}
  |\vor(\X(\alpha,t),t)| &= \vxi(\X(\alpha, t), t) \cdot \nabla_\alpha
  \X(\alpha, t) \cdot \vxi_0(\alpha) |\vor_0(\alpha)|\\ &=
  \frac{\partial s}{\partial \beta} |\vor_0(\alpha)|\,,
  \label{eq:t2stretch}
\end{align}
meaning that the local stretching of the length of a vortex line
segment that is transported with the flow is equivalent to the growth
of vorticity at the corresponding transported fluid element.

This result can be transformed into a bound for the length of a vortex
line by the vorticity along this line. Denote with $l(t)$ the length
of the vortex line segment $L_t$ at time $t$ and define with
\begin{equation}
  \Omega_L(t) := \|\vor(\cdot,t)\|_{L^\infty(L_t)}
\end{equation}
the maximum vorticity on the vortex line segment. Furthermore, let
\begin{equation}
  M(t) := \max(\|\nabla \cdot \vxi\|_{L^\infty(L_t)},\|\kappa\|_{L^\infty(L_t)})
\end{equation}
be the quantity of vortex line convergence $\nabla \cdot \vxi$ and
vortex line curvature $\kappa$, and define $\lambda(L_t) :=
M(t)l(t)$. Then, the relative increase of the length of the vortex
line segment in a time interval, $l(t)/l(0)$, is bounded as
\begin{equation}
  \label{eq:T2lemma}
  e^{-\lambda(L_t)} \frac{\Omega_l(t)}{\Omega_l(0)} \le \frac{l(t)}{l(0)} \le e^{\lambda(L_0)} \frac{\Omega_l(t)}{\Omega_l(0)}\;.
\end{equation}

This result, which is a slight modification of a lemma in
\cite{deng-hou-yu:2005}, may be viewed in its own right: The relative
increase in length along a time-interval is bounded by the vorticity
increase and a factor $\exp(\pm\lambda(L_t))$. Thus, $\lambda(L_t)$ is
a dimensionless number, characterizing the geometric ``tameness'' of
the vortex line filament.

As depicted in Fig. \ref{fig:T2lambda}, a vortex line segment has a
huge $\lambda(L_t)$, if its maximum curvature is large, relative to its
length (the segment is ``kinked'' instead of ``curved''), or if the
surrounding vortex lines collapse to the considered segment in at least
a point (the surrounding is ``tightening'' instead of ``parallel''). A
relatively unbent vortex line segment with approximately parallel
neighboring vortex lines possesses a low value of $\lambda(L_t)$. This
quantifies the constricted notion of ``relatively straight'' and
``smoothly directed'' given in \cite{constantin-fefferman-majda:1996}
in a sharper way.

\subsection{Lagrangian evolution of vortex line segments}

\begin{figure}[t]
  \centering
  \includegraphics[width=0.45\linewidth]{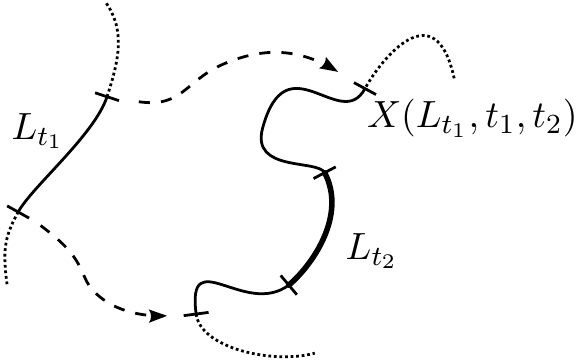}
  \caption{Visualization of the evolution of a vortex line segment in
    theorem 2: At a late time $t_2 > t_1$ the segment $L_{t_2}$ has to
    be included in the Lagrangian evolution of of $L_{t_1}$, but is
    free to be just a fraction of it. \label{fig:T2}}
\end{figure}
Connecting the stretching process stated above with the Lagrangian
accumulation of vorticity,
\begin{equation}
  \frac{D}{Dt} |\vor| = \left[(S \vxi) \cdot \vxi \right] |\vor|\;,
  \label{eq:lagrangevorticity}
\end{equation}
where $S = 1/2(\nabla \vec{u} + \nabla \vec{u}^T)$ is the strain
tensor, and noting that the \textit{curvature} $\kappa$ of the vortex
line $L_t$ fulfills
\begin{equation}
  \label{eq:defcurvature}
  \kappa \vec{n} = \frac{\partial \dot{L_t}(s)}{\partial s} = \frac{\partial \vxi}{\partial s} = (\vxi \cdot \nabla) \vxi\;,
\end{equation}
with $\vec{n} = \ddot{L}_t/|\ddot{L}_t|$ being the unit normal vector
of the vortex line, the Lagrangian evolution of vortex line
stretching, by inserting eq. (\ref{eq:t2stretch}), becomes
\begin{equation}
  \frac{D}{Dt} \left( \frac{\partial s}{\partial \beta} \right) =
    \frac{\partial}{\partial \beta}( \vel \cdot \vxi) - \kappa(\vel
    \cdot \vec{n}) \left( \frac{\partial s}{\partial \beta} \right)\;.
    \label{eq:lagstretching}
\end{equation}
At this point it becomes obvious how the process of vortex line
stretching interacts with the velocity in two distinct ways: The
velocity in direction of the vortex line elongates the segment by
drawing it out, while a part of the velocity normal to the vortex line
increases the segment's length by enlarging its curves.

Integrating (\ref{eq:lagstretching}) along the vortex line (from
$\beta_1$ to $\beta_2$) and over time (from $0$ to $t$) results in
\begin{equation}
  \label{eq:lengthT2}
  l(t) \le l(0) + \int_{0}^t \left[ U_\xi(\tau) + \lambda(\tau) U_n(\tau) \right] \mathrm{d}\tau\;,
\end{equation}
for
\begin{align*}
  U_\xi(t) &:= \max_{\x, \vec{y} \in L_t} |(\vel \cdot \vxi)(\x,t) - (\vel \cdot \vxi)(\vec{y},t)|\\
  U_n(t) &:= \max_{L_t}|\vel\cdot \vec{n}|
\end{align*}
Instead of starting the above reasoning at time $t=0$, the results are
identical for a later time $0 < t_1 < t$. This result may be
understood as an upper bound for vortex line stretching in terms of
velocity and vortex line geometry. In conjunction with the connection
between length increase and vorticity amplification, given in equation
(\ref{eq:T2lemma}), one arrives at
\begin{equation*}
  \Omega_l(t) \le \Omega_l(0) e^{\lambda(L_t)} \left[ 1 +
    \frac{1}{l(0)} \int_{0}^t (U_\xi(\tau) + \lambda(\tau)
    U_n(\tau)) \mathrm{d}\tau \right]\;.
\end{equation*}
This is an inequality for the control of growth rate of the vorticity
by geometric flow properties. From this estimate, by combining it with
BKM to distinguish critical from sub-critical vorticity growth, the
central non-blowup criterion of \cite{deng-hou-yu:2005} is derived:

\vspace{0.5ex}
\textbf{Deng-Hou-Yu theorem 2:} \textit{Assume there is a family of
  vortex line segments $L_t$ and $T_0 \in [0,T)$, such that $L_{t_2}
    \subseteq \X(L_{t_1}, t_1, t_2)$ for all $T_0 < t_1 < t_2 < T$. We
    also assume that $\Omega(t)$ is monotonically increasing and
    $\|\omega(t)\|_{L^\infty(L_t)} \geq c_0 \Omega(t)$ for some $c_0 >
    0$ when $t$ is sufficiently close to $T$. Furthermore, we assume
    that}
  \begin{enumerate}[(i)]
  \item $\lambda(L_t) \leq C_0,$
    \label{itm:T2_2}
  \item $l(t) \gtrsim (T-t)^B$ \textit{for some} $B \in (0,1)$.
    \label{itm:T2_3}
  \item $U_{\xi}(t) + U_n(t) \lambda(L_t) \lesssim (T-t)^{-A}$
    \textit{for some} $A < 1-B$
    \label{itm:T2_1}
  \end{enumerate}
  \textit{Then there will be no blowup in the 3D incompressible
  Euler flow up to time T.}
\vspace{0.5ex}

Here, $a(t) \lesssim b(t)$ means there exists a constant $c \in
\mathbb{R}$ such that $|a(t)|~<~c\,|b(t)|$ (and accordingly for
$a(t)~\gtrsim~b(t)$). The choice of Lagrangian vortex segments is
sketched in Fig. \ref{fig:T2}.

It should be noted that theorem 2 again includes assumptions on the
dimensionless number $\lambda(L_t)$. Especially assumption
(\ref{itm:T2_2}) poses a uniform bound in time for
$\lambda(L_t)$. This translates to words as the process of ``zooming
in'' to the location of maximum vorticity in order to keep the
considered vortex line segment relatively straight in comparison to
its length. The assumed accompanying collapse in length to keep
$\lambda(L_t)$ bounded is then linked in its growth rate to the blowup
of the velocity components.

It is worth mentioning that the above presented criterion, even though
it is obviously inspired by the classical geometric criteria, still
differs in crucial aspects. The posed assumptions are purely local and
restricted to the geometry of a single critical vortex line
filament. Assumptions on the velocity do not, in contrast to
Constantin et al. \cite{constantin-fefferman-majda:1996}, impose a
uniform bound (which is not observed in simulations), but allow for a
finite-time blowup of velocity, strictly connected in its growth rate
to the geometrical evolution of the filament. The vortex line segment
itself is not assumed to be of $O(1)$ length (as in
\cite{cordoba-fefferman:2001}) or to be contained in an $O(1)$-region
(which, again, was not observed in simulations). These aspects in
combination render it a promising theorem to be directly tested by
numerical simulations.

\subsection{Scenarios for finite-time singularities}

It has been established that a singularity of the Euler equations in
finite time necessitates rapid accumulation of vorticity. Locally,
vorticity-strain coupling is identified as the mechanism for nonlinear
amplification in finite time. Globally, the notion and possibility of
self-similar or locally self-similar collapse to a point is
introduced. These aspects serve as a basis for the construction of
initial conditions suitable for the possible formation of a
finite-time singularity.

Due to the incompressibility condition, the trace of $S$ vanishes and
due to the symmetry of $S$ its eigenvalues $\lambda_i$ are real and
the corresponding eigenvectors $\vec{v}_i$ are orthogonal. Thus
$\lambda_1$, the biggest eigenvalue, fulfills $\lambda_1 > 0$ in
regions with non-vanishing strain. As a direct consequence of equation
(\ref{eq:lagrangevorticity}), if $\lambda_1$ is proportional to the
vorticity in direction $\vec{v}_i$ then the vorticity growth would be
compatible with a finite-time singularity according to BKM. This
coupling of the strain to the vorticity is crucial. If the strain rate
is constant instead, the growth in vorticity is merely
exponential. Several cases have been suggested in which this mechanism
of coupling may take place.

It should be noted that, contrary to expectations, the
vorticity-strain coupling does not readily appear in nature. One would
expect a tendency of the vorticity to align itself to the eigenvector
of the strain tensor with the largest eigenvalue all by itself, since
the parallel component is amplified, while the orthogonal components
are damped or stay nearly constant. Nevertheless, for viscid turbulent
flows, quite a different behavior is observed both in numerical
simulations and experiments: The vorticity is most likely to align to
the intermediate eigenvector of the strain tensor
\cite{ashurst-kerstein-kerr-gibson:1987, meneveau:2011,
  chevillard-meneveau:2011} and similarly for the Euler equation
\cite{pumir-siggia:1990}. One can therefore expect that functional
vorticity-strain coupling is inherently unstable. The process has to
be designed ``artificially'' by choosing suitable initial conditions.

A more precise notion of the process of vorticity alignment in
turbulent flows is given by Hamlington, Schumacher and Dahm
\cite{hamlington-schumacher-dahm:2008}. They distinguish, evaluating
the Biot-Savart law numerically, between strain induced
\textit{locally} by the immediate neighborhood and \textit{globally}
by long-range interaction. For turbulent flows, they observe a most
likely alignment of the vorticity to the most positive eigenvector of
the global strain. Taking into account also the local strain restores
the alignment of vorticity to the intermediate eigenvector. Contrary
to this, for the successful emergence of a finite-time singularity of
the Euler equations in a point-wise sense, vorticity-strain coupling
should be induced by the local strain to support a collapse to a
point. A numerical application of this technique from
\cite{hamlington-schumacher-dahm:2008} to Euler blowup simulations
will be presented elsewhere.

\section{Initial conditions}

Different initial conditions were introduced and subsequently improved
or refined to construct flows with prolonged intervals of
vorticity-strain coupling, and it seems natural to search for
techniques to artificially keep the coupling existent. One such
technique is the introduction of symmetries to the flow. Early
examples such as the Taylor-Green vortex \cite{taylor-green:1937} or
Kerr's initial conditions \cite{kerr:1993, kerr:2005, hou-li:2006} are
employing such symmetries.

\subsection{Reflectional symmetries}
\label{ssec:reflectionalsymmetries}

\begin{figure}[t]
  \centering
  \includegraphics[width=0.5\textwidth]{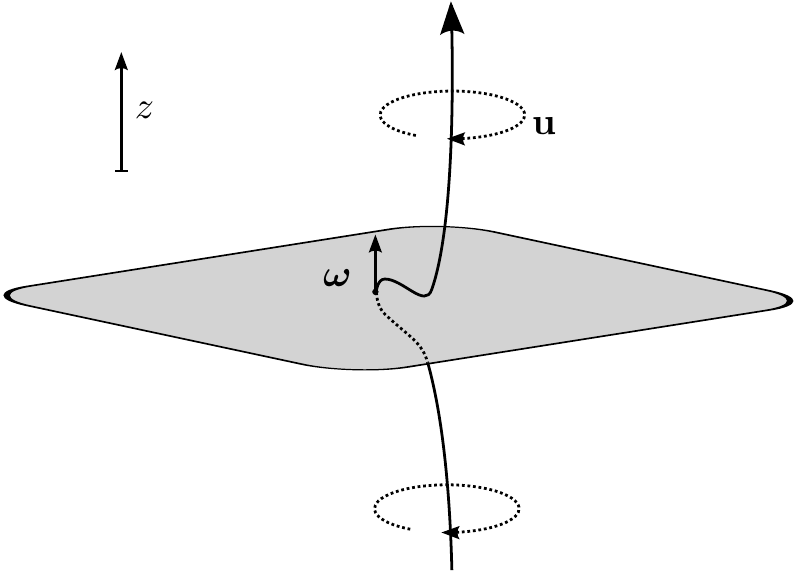}
  \caption[Reflectional symmetries and intersecting vortex
    tubes]{Effects of planes of reflectional symmetry on intersecting
    vortex tubes: In the symmetry plane, the vorticity $\vor$ is
    normal and the strain tensor possesses a parallel eigenvector with
    corresponding eigenvalue $S_{zz}$. The curvature $\kappa$ in the
    symmetry plane has to increase in order to support critical
    vorticity-strain coupling.}
  \label{fig:reflectionalsymmetry}
\end{figure}

Consider the plane $z=0$ to be a plane of reflectional symmetry, as
shown in Fig. \ref{fig:reflectionalsymmetry}, defined by
\begin{align*}
  u_x(x,y,z) &= u_x(x,y,-z)\\
  u_y(x,y,z) &= u_y(x,y,-z)\\
  u_z(x,y,z) &= -u_z(x,y,-z)
\end{align*}
for the velocity vector field, which leads to $u_z = 0$ in the plane
of symmetry. Accordingly, the vorticity obeys
\begin{align*}
  \omega_x(x,y,z) &= - \omega_x(x,y,-z)\\
  \omega_y(x,y,z) &= - \omega_y(x,y,-z)\\
  \omega_z(x,y,z) &= \omega_z(x,y,-z)
\end{align*}
and in particular $\omega_x = \omega_y = 0$ or $\vor = \omega_z
\hat{e}_z$ in the plane of symmetry. Due to these properties, the
strain tensor has $S_{xz} = S_{yz} = S_{zx} = S_{zy} = 0$, or
\begin{equation}
  S = \begin{pmatrix} S_{xx}&S_{xy}&0\\ S_{xy}&S_{yy}&0 \\ 0&0&S_{zz} \end{pmatrix}\,.
\end{equation}
It immediately follows that the eigenvector corresponding to the
eigenvalue $S_{zz}$ is directed normally to the symmetry plane, and the
vorticity vector is aligned to it. Note that this is the sole
consequence of the reflectional symmetry and is in no way influenced
by the flow. A vortex tube normal to the symmetry plane therefore
seems like a natural candidate for critical accumulation of vorticity
by means of vorticity-strain coupling: All that is needed is a
sufficiently long period of time in which $S_{zz}~\sim~\omega_z$ at
one point of the symmetry plane.

This possibility has been analyzed by Pelz \cite{pelz:2001}: Taking
into account only the $zz$-component, the strain tensor in the plane
of symmetry is given by the Biot-Savart law as
\begin{equation}
  \label{eq:symmetrystrain}
  S_{zz} = \frac{3}{4\pi} \int \left( (x-x') \omega_y(\x') - (y-y') \omega_x(\x') \right)\frac{(z-z')}{|\x-\x'|^5} \mathrm{d}\x'\;.
\end{equation}
Equation (\ref{eq:symmetrystrain}) shows that $S_{zz}$ in the plane of
symmetry does not scale with $\omega_z$, but does instead depend on
$\omega_x$ and $\omega_y$, which are both equal to zero in the $z=0$
plane. Yet, in close proximity to the plane, $\omega_x$ and $\omega_y$
may grow, depending on the curvature of the vortex line intersecting
the symmetry plane: A Taylor expansion around the $z=0$ plane yields:
\begin{align*}
  \omega_i &= h \left. \frac{\partial \omega_i}{\partial z} \right|_{z=0} = h \kappa_i \omega_z(z=0)
\end{align*}
for small $h$ up to first order, for $\kappa_i = \kappa
\vec{n}_i$. Therefore, if the curvature is huge close to the plane of
symmetry, $\omega_i$ with $i \in \{x,y\}$ approximately scales like
$\omega_z$ and thus $S_{zz}$ may scale with $\omega_z$ too. However,
for this to happen, we need $\kappa \approx 1/h$. As a matter of fact,
the dimensionless number $\kappa_i h$ plays a similar role as the
characteristic geometric number $\lambda(t)$ introduced in section
\ref{ssec:vortexlinestretching}. For $S_{zz}$ to blow up like
$\omega_z$, the curvature has to increase in a way to counter the
shrinking of the length scale $h$.

On the other hand, the axial strain $S_{zz}$ stretches the vortex tube
in $z$-direction. This counteracts any increase in curvature to a
certain degree. More precisely, the Lagrangian evolution of the
curvature components $\kappa_x$ and $\kappa_y$ were calculated in
\cite{pelz:2001}:
\begin{align*}
  \frac{D}{Dt} \kappa_x &= (S_{xx} - 2 S_{zz}) \kappa_x + S_{yy} \kappa_y + \partial_z S_{yz}\\
  \frac{D}{Dt} \kappa_y &= S_{xx} \kappa_x + (S_{yy} - 2 S_{zz}) \kappa_y + \partial_z S_{xz}\;.
\end{align*}
The axial strain $S_{zz}$ diminishes both $\kappa_x$ and $\kappa_y$.

These counteracting processes of vortex line geometry are by no means
analytically exact, since all long-range interactions have been
ignored. Nevertheless, they constitute an intrinsic resistance of a
single vortex line to ``self-stretch'' in a critical way. The argument
may be readily translated to the case of (perturbed) anti-parallel
vortex tubes: Since no other components of vorticity are introduced,
$S_{zz}$ still only depends on $\omega_x$ and $\omega_y$ which in turn
rely on high curvature to scale like $\omega_z$ close to the plane of
symmetry.

One way to counter this is to induce the axial strain by neighboring
tubes instead of relying on a sufficiently large kink. This will be
presented in the following section by introducing additional
rotational symmetry.

\subsection{High symmetry initial conditions}

One notable high-symmetry flow was introduced by Kida \cite{kida:1985}
and has subsequently been used extensively to probe a possible Euler
blowup numerically \cite{boratav-pelz:1994b, pelz:2003,
  cichowlas-brachet:2005} or analytically
(e.g. \cite{ng-bhattacharjee:1996}) as well as study the onset of
turbulence (e.g. \cite{boratav-pelz:1994}). The Kida-Pelz flow has a
three-fold rotational symmetry about the diagonal and a reflectional
symmetry about all three Cartesian planes. Flows with these two
properties are termed as invariant under the full octahedral group
\cite{pelz:2001}. The Euler (and Navier-Stokes) equations preserve the
Kida-Pelz symmetries. In general one can write these initial
conditions as
\begin{align}
  \label{eq:kidasymm}
  v(x,y,z) &= \sum_{l,m,n} a_{lmn} \sin(lx) \cos(my) \cos(nz)\\
  \vec{u} &= (u_x, u_y, u_z)^T = (v(x,y,z), v(y,z,x), v(z,x,y))\;.
\end{align}
This means, for a computational domain spanning the interval
$[0,~\pi]$ in all three dimensions, that the normal component of the
velocity field is anti-symmetric under reflection at the Cartesian
planes while the tangential components are symmetric. Combining this
with the three-fold rotational symmetry adds up to a total memory
saving factor of $1/24$.

On the same time there is reason to hope that these rather artificial
symmetries encourage singular behavior if the initial conditions are
constructed accordingly. When assuming a localized vortex tube
intersecting the symmetry plane normally, as depicted in Fig.
\ref{fig:initialgrauer}, its mirror images result in a total of six
pairs of anti-parallel vortex tubes. It has been proposed by Pelz
\cite{pelz:2001} that the strain induced by the rotational images of
each tube, assuming a velocity field supporting a collapse to the
origin, may lead to the desired vorticity-strain coupling without
being subject to the counteraction of strain and curvature in the
planes of symmetry. Provided that the vortex dodecapole retains its
shape during collapse, this scenario could lead to a finite-time
collapse to the origin.

\subsubsection{Vortex dodecapole initial conditions}

\begin{figure}[t]
  \centering
  \includegraphics[width=0.45\textwidth]{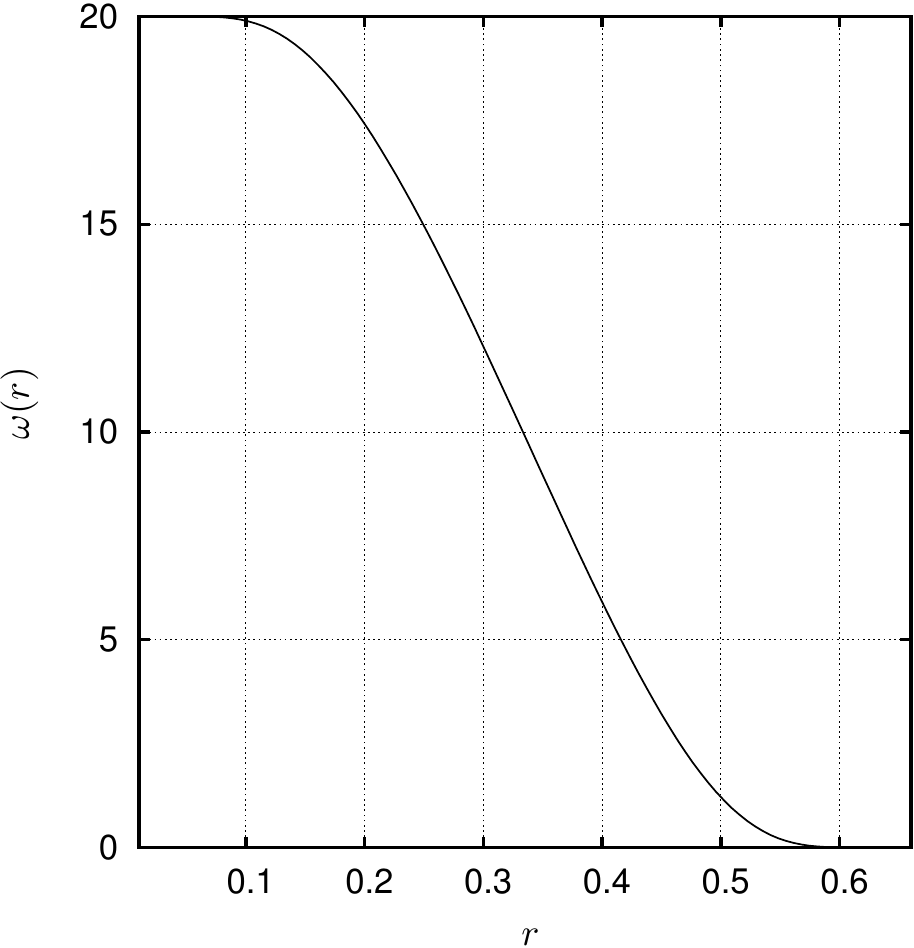}
  \includegraphics[width=0.45\textwidth]{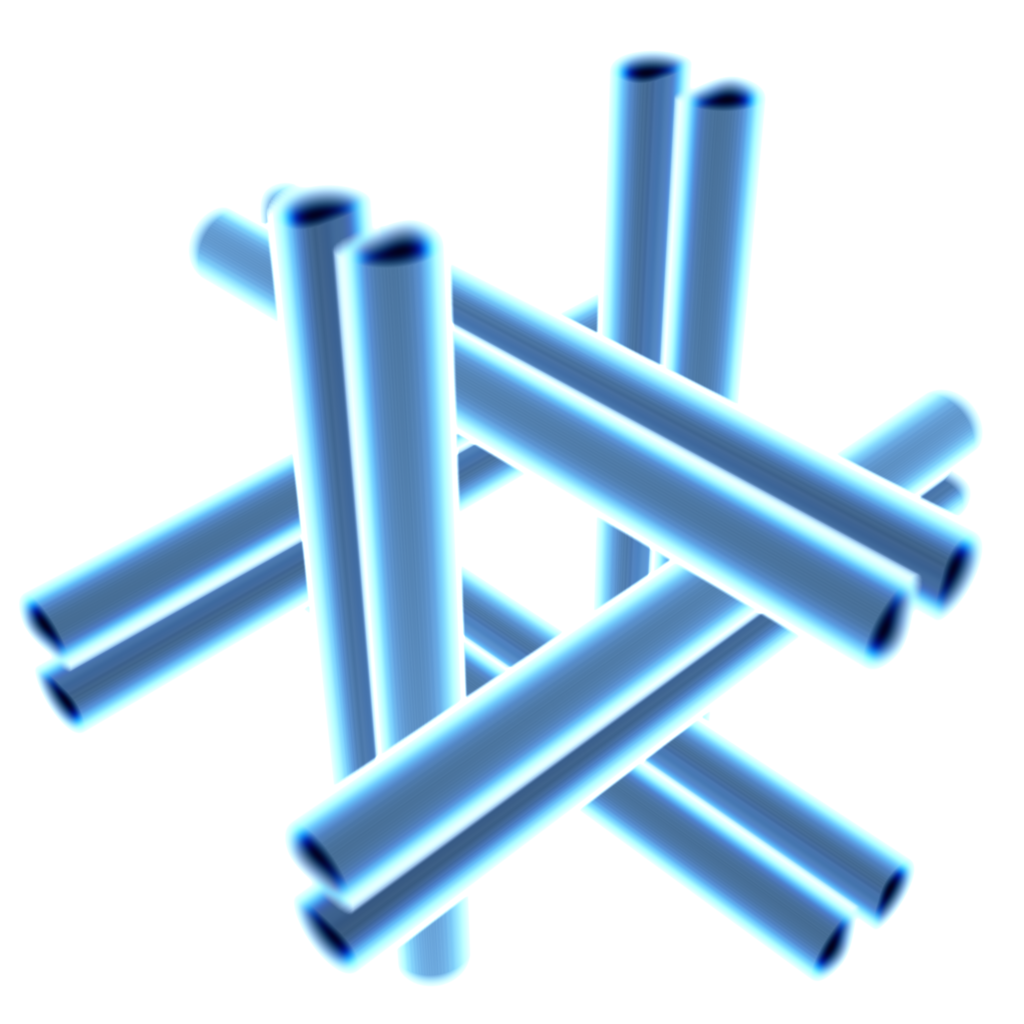}
  \caption{Left: Vorticity profile of one tube of the
    12-tube initial condition (with $A=20$). Right:
    Volume plot of the vorticity for the whole domain.}
  \label{fig:initialgrauer}
\end{figure}

One form of these initial conditions is based on the idea to already
start with six dipoles consisting of vortices of a designated
vorticity profile. An example is the vortex dodecapole initial
condition \cite{grafke-homann-dreher-grauer:2008} with a vorticity
profile given by
\begin{equation}
\label{eq:initialgrauer}
\omega(r) = \left \lbrace 
\begin{aligned} &
  -A\left[1-\exp\left(-e^2\log(2)\frac{1}{3r}\exp\left(\frac{1}{\frac{3r}{2} - 1}\right)\right)\right] & \text{for } r<\frac{2}{3}\\
  & 0  & \text{for } r\ge\frac{2}{3}\\
\end{aligned} \right.
\end{equation}
where $r$ denotes the distance to the tube's center line. The
vorticity decreases with increasing distance to the center line and is
strictly zero for $r>2/3$. Thus, the vortex has compact support in the
$r$-$\varphi$-plane while still being smooth. Fig.
\ref{fig:initialgrauer} (left) displays the vorticity profile given
above, Fig. \ref{fig:initialgrauer} (right) shows the whole
dodecapole. Only one octant, i.e. three vortices, are simulated due to
symmetry. For the reasons lined out in the previous sections, this
kind of dodecapole appears to very promising in terms of developing a
finite-time singularity in the origin because of reciprocal strain of
the mirror tubes. It is furthermore susceptible to the analysis by the
presented geometric blowup criteria. Most of the diagnostics in this
paper are therefore performed on flows which are based on initial
conditions of this type.

\subsubsection{Lamb-dodecapole initial conditions}

\begin{figure}[t]
  \centering
  \includegraphics[width=0.49\textwidth]{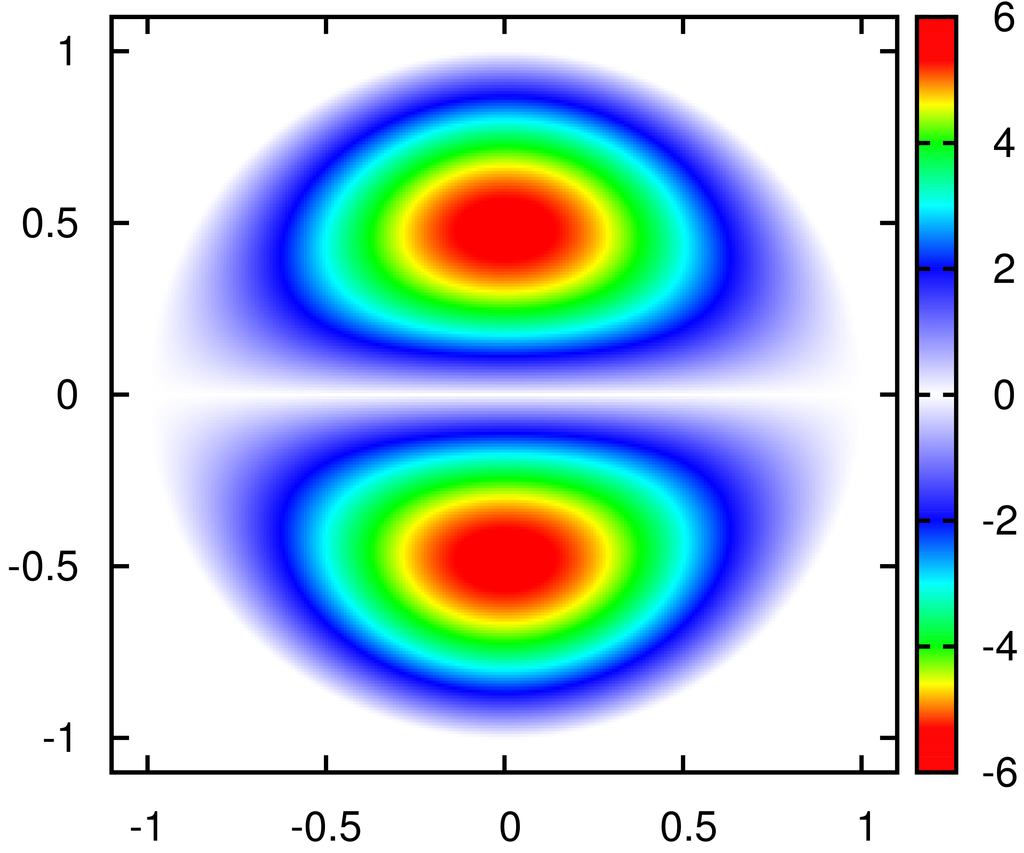}
  \includegraphics[width=0.49\textwidth]{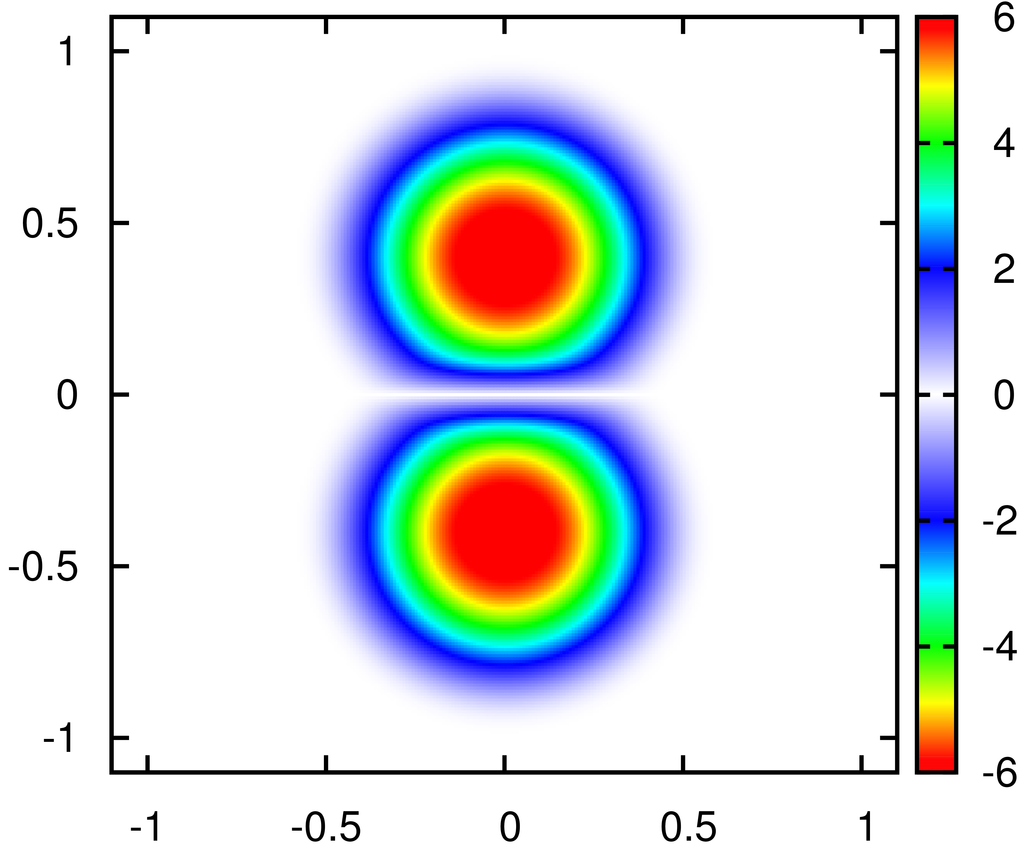}
  \caption{Comparison of vorticity profile for a Lamb dipole to the
    simple dipole. Left: Lamb dipole used in the
    Lamb-dodecapole initial conditions. Right: Dipole used in
    vortex dodecapole initial conditions. Both are scaled to fit in
    amplitude and size.}
  \label{fig:lambvsgrauer}
\end{figure}

For the Euler equations, a single stretch-free axisymmetric vortex may
have \textit{arbitrary} radial dependence for the vorticity to remain
stationary in time. The same is not true for vortex dipoles: An
isolated vortex dipole propagating through the domain does not
preserve its shape. This so-called \textit{vortex shedding} is
believed to influence and possibly suppress a self-amplifying
behavior \cite{orlandi-carnevale:2007}.

There are exact form-preserving dipole solutions of the 2-dimensional
Euler equations which may be used to construct initial conditions that
do not exhibit vortex shedding. The most famous is the Lamb-dipole
introduced by Lamb \cite{lamb:1932}. Following \cite{wu-ma-zhou:2006} it is
defined by
\begin{equation}
\label{eq:initiallamb}
\omega(r) = \left \lbrace 
\begin{aligned} &
  2Uk\frac{J_1(kr)}{J_0(ka)}\sin(\theta) & \text{for } r<a\\
  & 0  & \text{for } r\ge0\\
\end{aligned} \right.
\end{equation}
where $k$ is chosen such that $ka$ is the first zero of $J_1$,
i.e. $ka \approx 3.8317$. In Fig. \ref{fig:lambvsgrauer} this
vorticity distribution is compared to the vorticity profile given in
equation (\ref{eq:initialgrauer}). Note that even though the
distribution of vorticity for the Lamb dipole appears to be less sharp
than for the simple profile, it is not differentiable at $r=a$, as can
be seen in equation (\ref{eq:initiallamb}). Because of this, strictly
speaking, the Lamb dipole is an improper candidate for the search for
finite-time singularities. This issue is usually overcome by smoothing
high frequency components in order to smear out the discontinuity in
the gradient of the vorticity.

Orlandi and Carnevale \cite{orlandi-carnevale:2007} where the first to
use Lamb dipoles to construct a colliding pair of dipoles, observing a
rapid amplification of vorticity for a period of time, with a slowing
growth at later times due to either depletion of nonlinearity
\cite{frisch-matsumoto-bec:2003} or lack of resolution. The Lamb
dipole is used in a similar manner in the context of this paper to
form a Lamb dodecapole analogous to vortex dodecapole initial
conditions presented above.

\section{Numerical experiment}

Along the lines of the above presented mechanism for a finite-time
singularity for the Euler equations, a number of numerical simulations
were performed in the last decade to act as evidence for or against a
blowup. Beginning in the early 80s of the last century, numerous
numerical simulations with a variety of methods and schemes and
differing initial conditions where tested, from Pad\'e approximants
\cite{morf-orszag-frisch:1980}, vortex-segment methods
\cite{chorin:1981, chorin:1982} and vortex filament models
\cite{siggia:1985}, to projection methods \cite{bell-marcus:1992}, pseudo
spectral simulations \cite{brachet-meiron-orszag-nickel-etal:1983,
  brachet-meneguzzi-vincent-politano-etal:1992,
  cichowlas-brachet:2005, hou-li:2006}, Chebychew codes
\cite{kerr:1993} and adaptive mesh refinement \cite{pumir-siggia:1990,
  grauer-marliani-germaschewski:1998}. Despite ever growing
resolution, from $128^3$ \cite{bell-marcus:1992} up to $4096^3$
mesh-points of adaptive simulations
\cite{grafke-homann-dreher-grauer:2008} or latest pseudo-spectral
codes \cite{bustamante-brachet:2011}, results are often inconclusive
or even conflicting.

In this paper The need for high resolution is met via massively
parallel adaptive mesh refinement for the results presented. This high
resolution data is then analyzed on the basis of the geometric blowup
criteria presented in section \ref{sec:geometric}.

\subsection{Computational Framework}

All numerical simulations throughout this work were conducted using
the recently developed framework \textit{racoon~III} (refined adaptive
computations with object-oriented numerics, based on
\cite{dreher-grauer:2005}). Its key feature is the integration of
partial differential equations on adaptive grids on massively parallel
distributed computers. Most Euler blowup scenarios feature extremely
localized structures with steep gradients, where a fixed mesh would
under-resolve the crucial parts while wasting resources on the less
important ones.

Adaptive mesh refinement increases the locally available resolution,
but comes at the cost of additional computational overhead. It
complicates the framework in several ways. Most importantly, it
restricts the choice of numerical schemes to comparatively simple low
order finite difference or finite volume variants. A direct comparison
to high accuracy pseudo-spectral simulation was made in
\cite{grafke-homann-dreher-grauer:2008} for the case of Euler
equations. It was found that a resolution approximately 1.3 times
higher is needed to reach a comparable accuracy for the adaptively
refined code.

\begin{figure}
\begin{center}
  \includegraphics[angle=-90,scale=-1,width=0.95\textwidth]{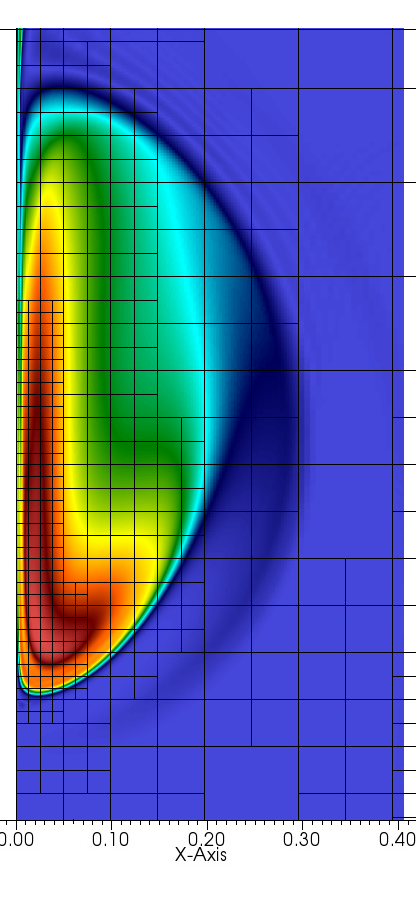}
  \caption[Refinement criterion for the simulation of the Euler
    equation with \textit{racoon~III}]{Refinement criterion for the
    simulation of the Euler equations with \textit{racoon
      III}. Regions with a large value for $\|\nabla \vec{u}\|$ are
    resolved higher. Each square represents a block with $16^3$
    cells. Shown is the absolute vorticity for a cross-section of one
    vortex tube.}
  \label{fig:wirbelamr}
\end{center}
\end{figure}

To decide which regions are to be refined, the norm of the gradient of
velocity, $\|\nabla \vel(\x,t)\|$ is compared to a threshold. If the
block is flagged as being under-resolved, it is bisected into $2^d$
child blocks that are redistributed among the available nodes. The
resolution of the parent block is thus effectively doubled. The
opposite happens for blocks that are over-resolved: $2^d$ blocks are
merged into one, the resolution at this location is halved. With this
procedure, the grid is constantly changing and adapting to the
simulation, as shown in Fig. \ref{fig:wirbelamr}, allowing high
resolution at critical locations but not wasting any resources for the
rest.

\begin{figure}
\begin{center}
  \includegraphics[width=0.45\textwidth]{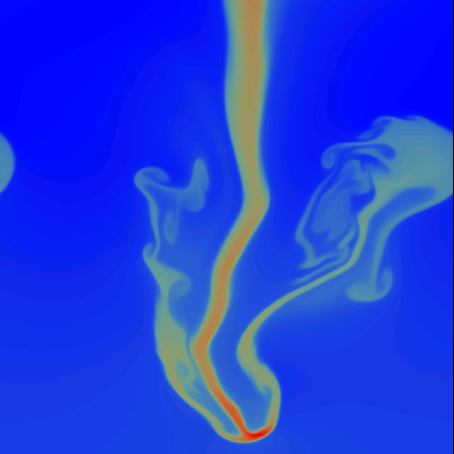}\hspace{0.05\textwidth}
  \includegraphics[width=0.45\textwidth]{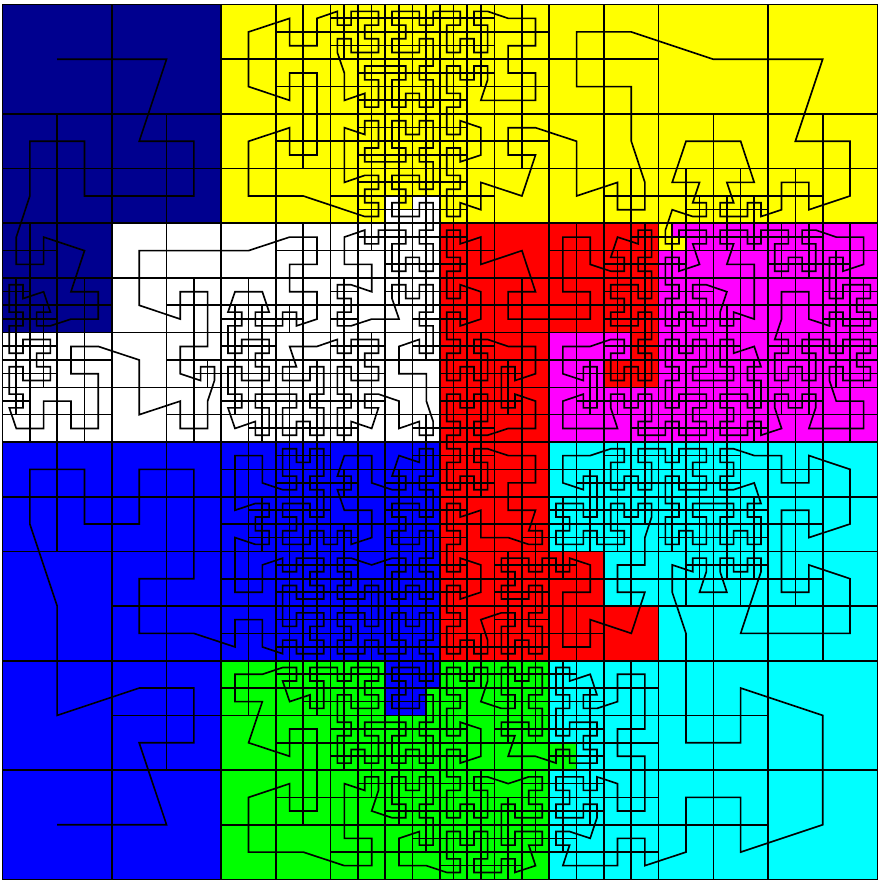}
  \caption[Adaptive mesh refinement and work balancing along a Hilbert
    curve]{Adaptive mesh refinement and dynamic load balancing. The
    workload is distributed among different processors along a
    space-filling Hilbert curve.}
  \label{fig:Namr}
\end{center}
\end{figure}
Since communication between different nodes is the smallest bottleneck
due to limited bandwidth and high latency, it is advantageous to
arrange the blocks in a way that physically close blocks are located
on the same node. Even if this seems to be pretty straight-forward for
normal grids, it poses a larger problem for adaptive grids with
different resolutions. In \textit {racoon~III}, blocks are distributed
along a space-filling Hilbert curve, as sketched in Fig.
\ref{fig:Namr}.

This ensures that proximate blocks are located on the
same node even if the grid is not fixed. Currently \textit{racoon~III}
uses a slightly different approach, using independent Hilbert curves
for each level, since inter-level communication is the most frequent
type of communication for common problems. Every time the grid changes
when adapting to the current situation, the Hilbert curve is
recalculated, as is the workload for each node. If an imbalance is
detected, the blocks are redistributed along the curve, each node
getting roughly the same amount of blocks.

The numerical scheme consists of a strong stability preserving third
order Runge-Kutta \cite{shu-osher:1988} time integrator combined with
a third order shock-capturing CWENO scheme \cite{kurganov-levy:2000}
to reduce oscillations in the presence of strong gradients. The
integrated equation is the vorticity formulation of the Euler
equations,
\begin{equation}
  \label{eq:vorticity}
  \frac{\partial}{\partial t} \vor + \nabla \times \left( \nabla (\vel \otimes \vel) \right) = 0\;,
\end{equation}
employing a vector potential formulation $\Delta \vec{A} = -\vor$ with
$\vel = \nabla \times \vec{A}$ to ensure solenoidality of the
vorticity vector field $\vor$. The associated Poisson equation is
solved with a second order parallel and adaptive multigrid algorithm
via a full approximation scheme to account for the non-uniform meshes,
with flux correction at refinement interfaces. Interpolation on the
coarse-fine interfaces is done in $\vor$ for the whole scheme to
ensure the highest possible accuracy in the critical variable. Passive
tracer particles are injected into the flow for the tracking of
Lagrangian vortex line segments. The above third order Runge-Kutta is
also used for the time integration of the tracer particles and the
space integration of vortex lines.

\begin{figure}
\begin{center}
  \includegraphics[width=0.9\textwidth]{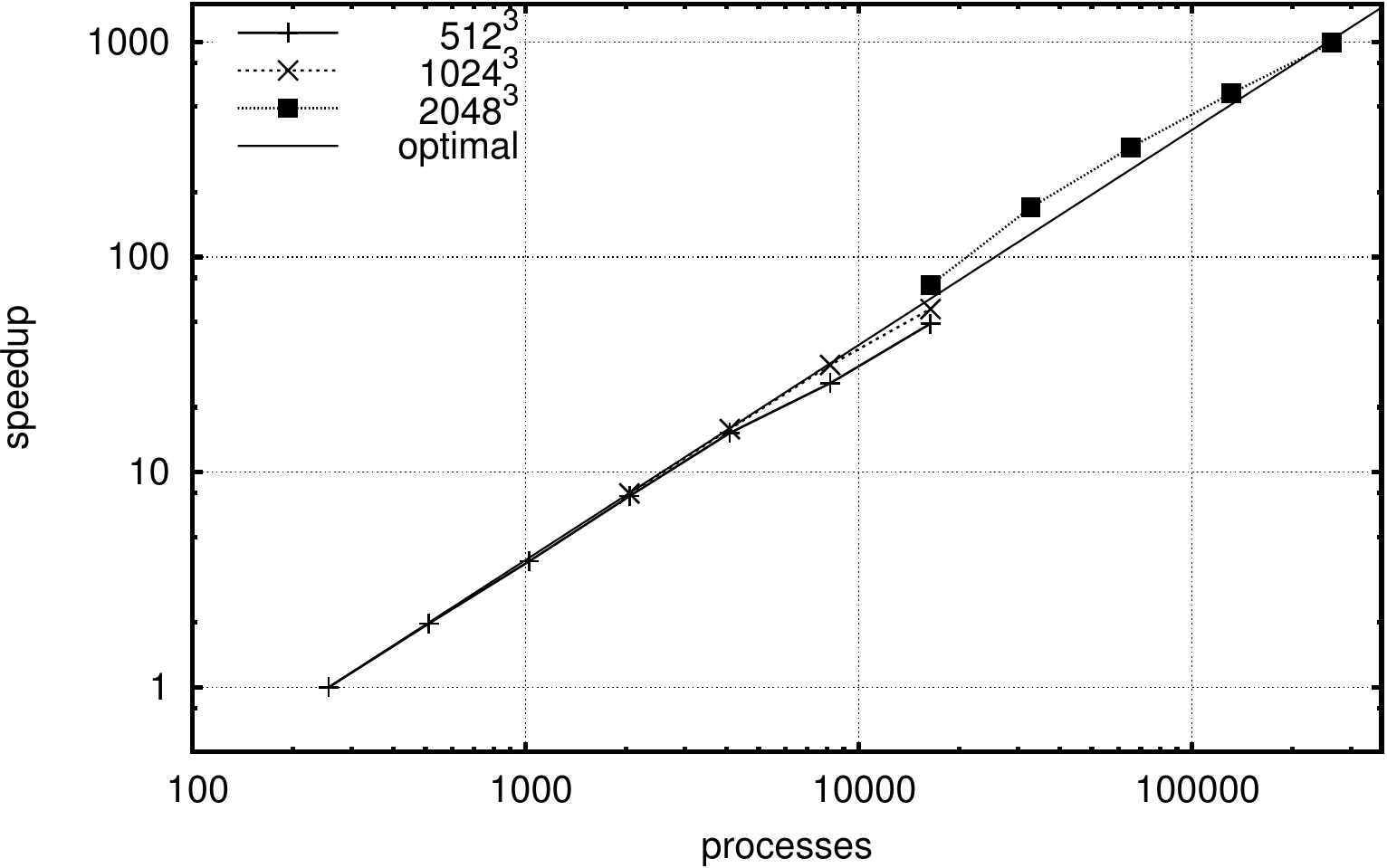}
  \caption[Scaling properties of the framework \textit{racoon}]{Mixed
    weak and hard scaling for the framework \textit{racoon} for a
    hyperbolic test problem. The scaling is close to linear for up to
    262144 cores.}
  \label{fig:racoonscaling}
\end{center}
\end{figure}

The overall scaling for \textit{racoon} is depicted in Fig.
\ref{fig:racoonscaling} for a hyperbolic test prolem (compressible
MHD. It was measured on the BlueGene/P machine at Forschungszentrum
J\"ulich with a total number of 294912 cores. For a combination of
weak and hard scaling, the performance is close to linear up to 262144
cores, the maximum number tested. The elliptical problems encountered
when simulating the Euler equations (velocity projection or
calculation of the vector potential) are more difficult to parallelize
than the hyperbolic advection term due to their inherent non-local
nature. For each timestep, information travels only fractions of the
grid spacing in the advection step, but through the whole domain when
enforcing the incompressibility. This behavior is necessarily
reflected by the demands on communication between processes in
massively parallel simulations. With inclusion of the multigrid
algorithm, the scaling is efficient only up to 131072 cores.

\subsection{Evolution of the flow}

This section is devoted to the visible results of the actual
simulation of the presented vortex dodecapole configurations. For this
purpose, the CWENO vector potential formulation is used in conjunction
with adaptively refined meshes for simulations with a resolutions of
up to $8192^3$ effective grid points, taking into account the increase
in resolution due to the high symmetry of the initial conditions. Both
the Lamb-dodecapole and the vortex dodecapole are used as initial
conditions.

\subsubsection{Vortex dodecapole}

\begin{figure}[p]
  \centering
  \includegraphics[width=0.49\textwidth]{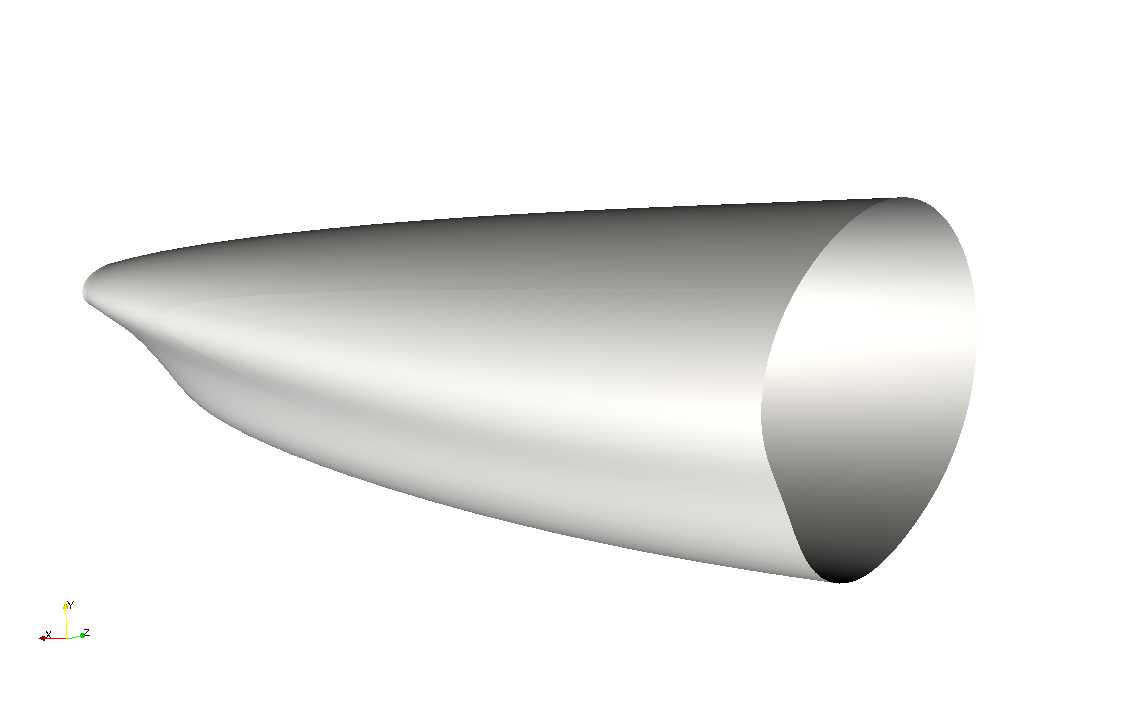}
  \includegraphics[width=0.49\textwidth]{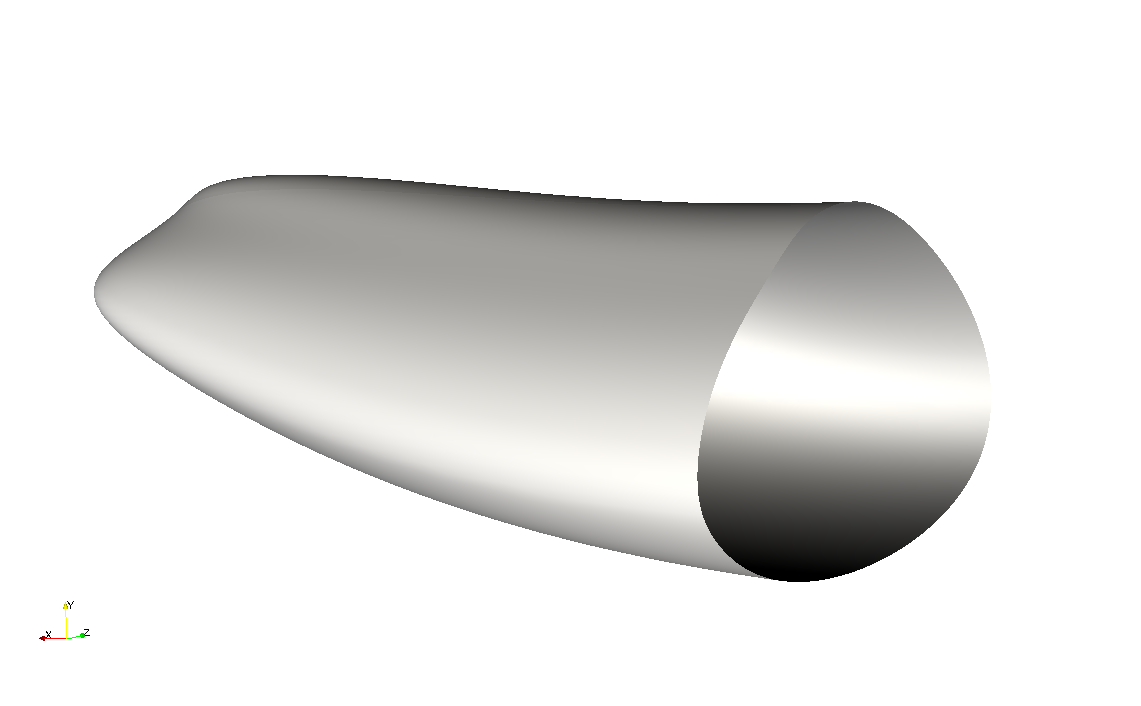}\\[2em]
  \includegraphics[width=0.49\textwidth]{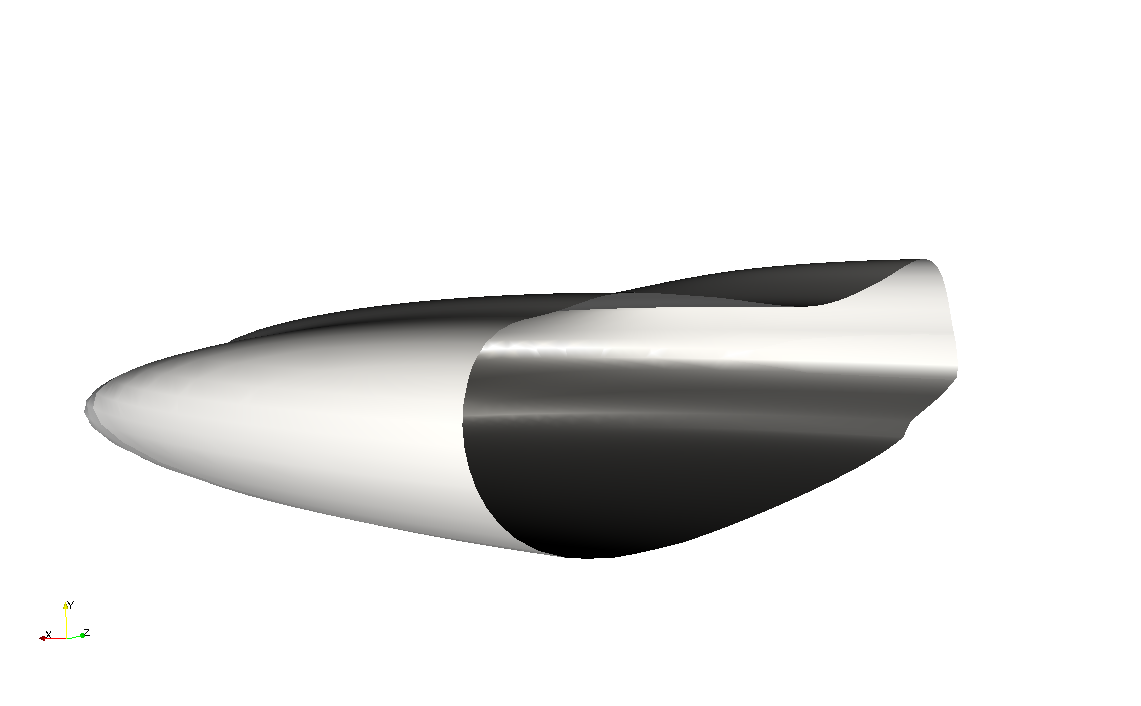}
  \includegraphics[width=0.49\textwidth]{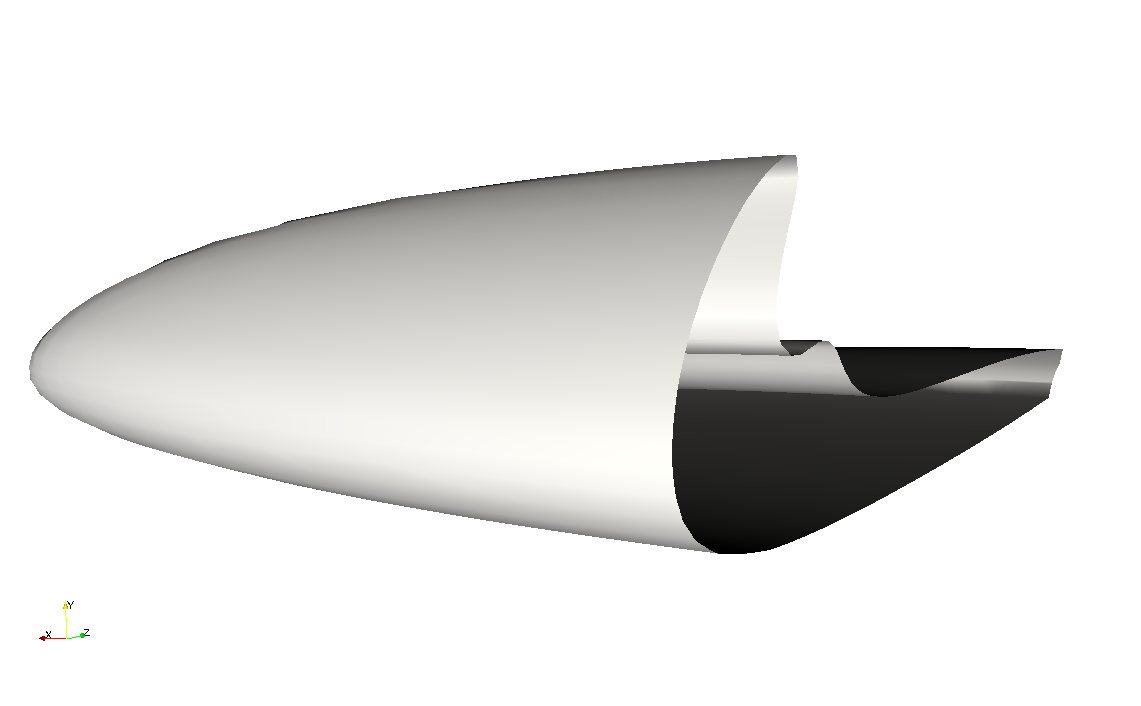}\\[4em]
  \includegraphics[width=0.49\textwidth]{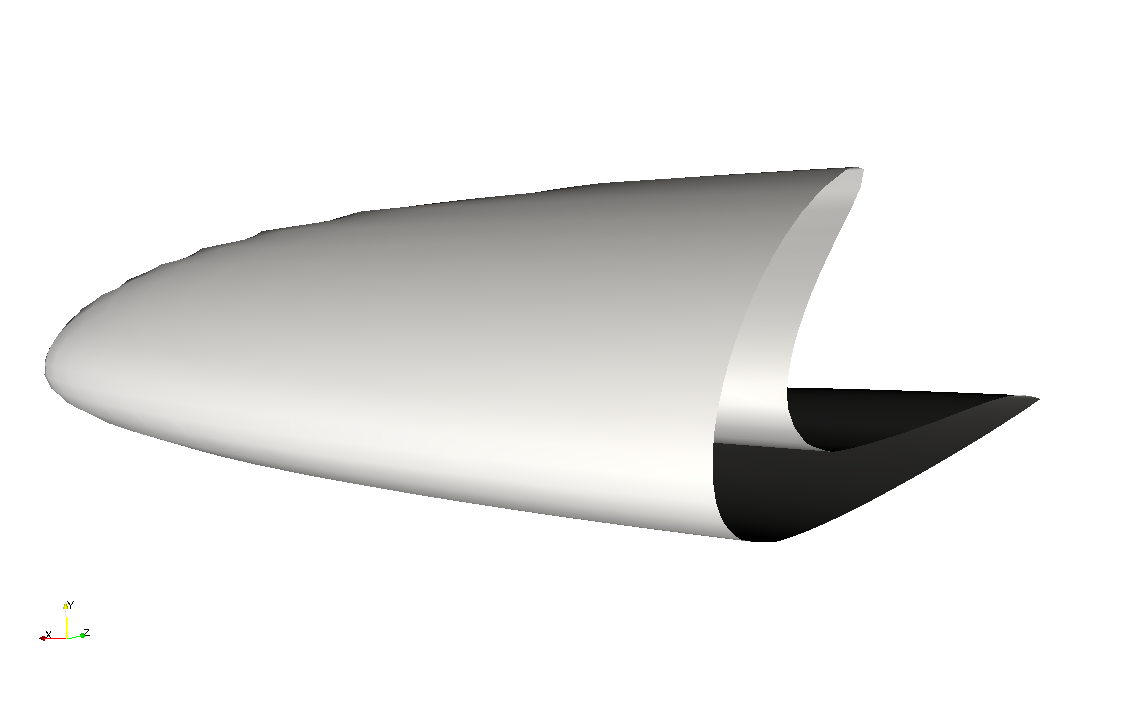}
  \includegraphics[width=0.49\textwidth]{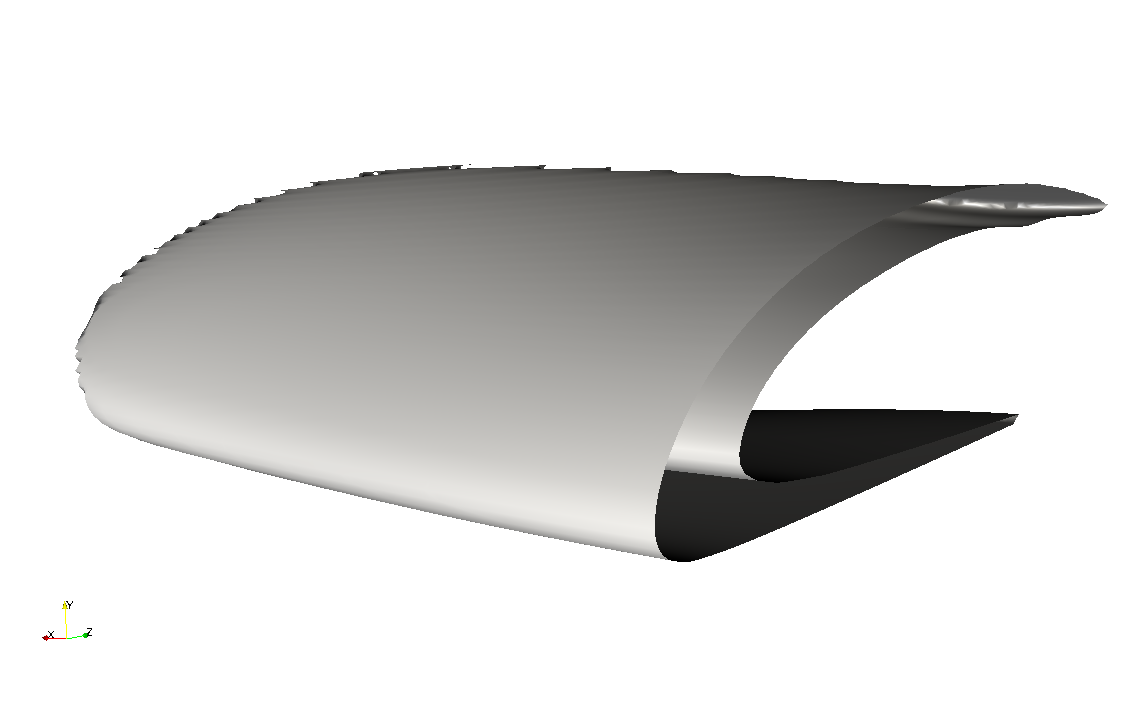}\\[2em]
  \caption{Evolution of the vortex dodecapole. Pictured are
    isosurfaces of the absolute vorticity, $|\vor(x,t)|$ at 75\% of
    the peak vorticity. Only one of twelve tubes is shown. The
    flattening of the vortex tube is followed by a roll-up. The
    developing secondary sheet finally exceeds the original sheet in
    size. All pictures are from run \texttt{amr1}.}
  \label{fig:evolutiongrauer}
\end{figure}
The vortex dodecapole was chosen as a prototype for the class of
vortex dodecapole initial conditions. Its main features are a smooth
vorticity profile with compact support and straight, unperturbed
initial vortex tubes.

Pictured in Fig. \ref{fig:evolutiongrauer} is the evolution in time
for the vortex dodecapole initial conditions. Shown are isosurfaces of
the absolute vorticity $|\vor(\x,t)|$ at 75\% of the peak vorticity
for different times. Due to the high symmetry, only one octant of the
computational domain is simulated. The figures therefore depict only
one half of a vortex tube, with twelve similar tubes in the total
domain. The initial phase of the development is depicted in the first
two sub-figures: The initially straight tube gets slightly stretched
due to interaction with the neighboring tubes. In the third frame, the
well-known flattening is in progress. The last three pictures present
the final stage of the flow, where the tip of the sheet rolls up and
forms a secondary vortex sheet. In the final figure, the secondary
sheet exceeds the original sheet in length. Its tip gets drawn out of
the collapsing region.

The appearance of the roll-up and the secondary vortex sheet are a
first evidence against a locally self-similar amplification and
collapse to a point: The initially round vortex tubes are severely
deformed and do not resemble their initial configuration in
shape. Furthermore, the possibility that the formation of a roll-up
may lead to the emergence of a tube-like structure which again form a
dodecapole arrangement is clearly conflicting the numerical evidence.
\begin{figure}[p]
  \centering
  \includegraphics[width=0.49\textwidth]{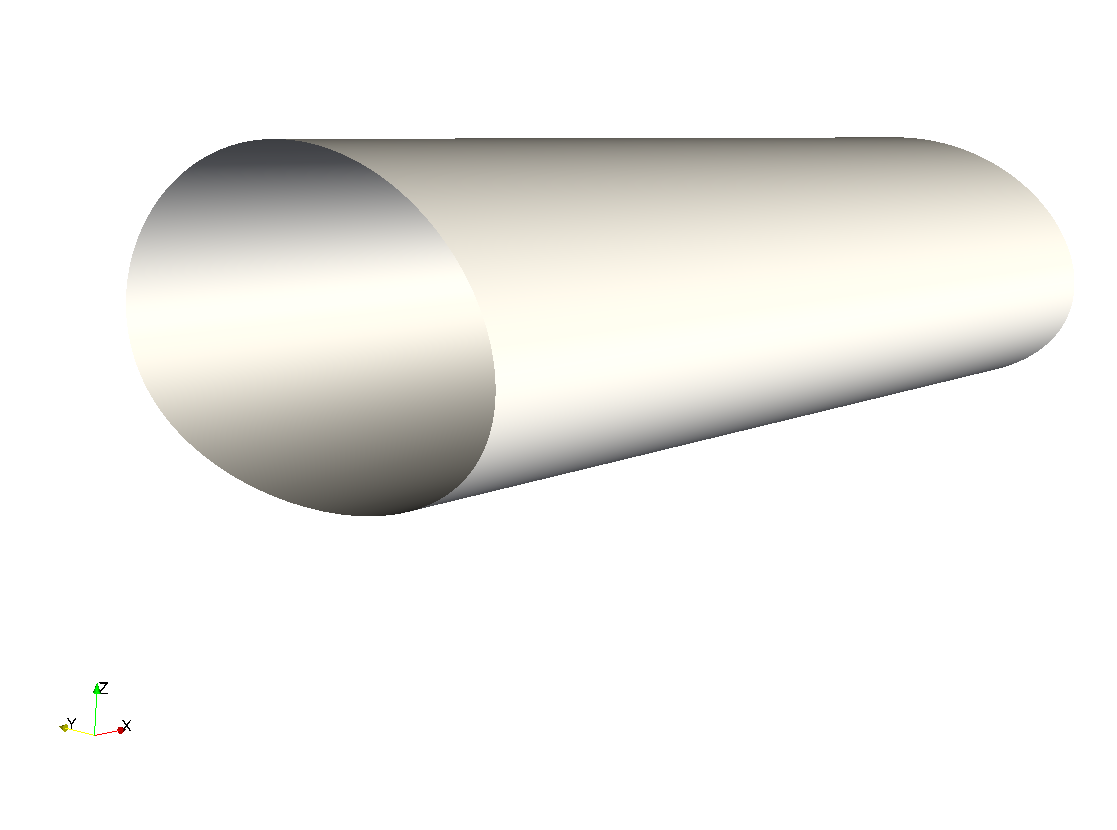}
  \includegraphics[width=0.49\textwidth]{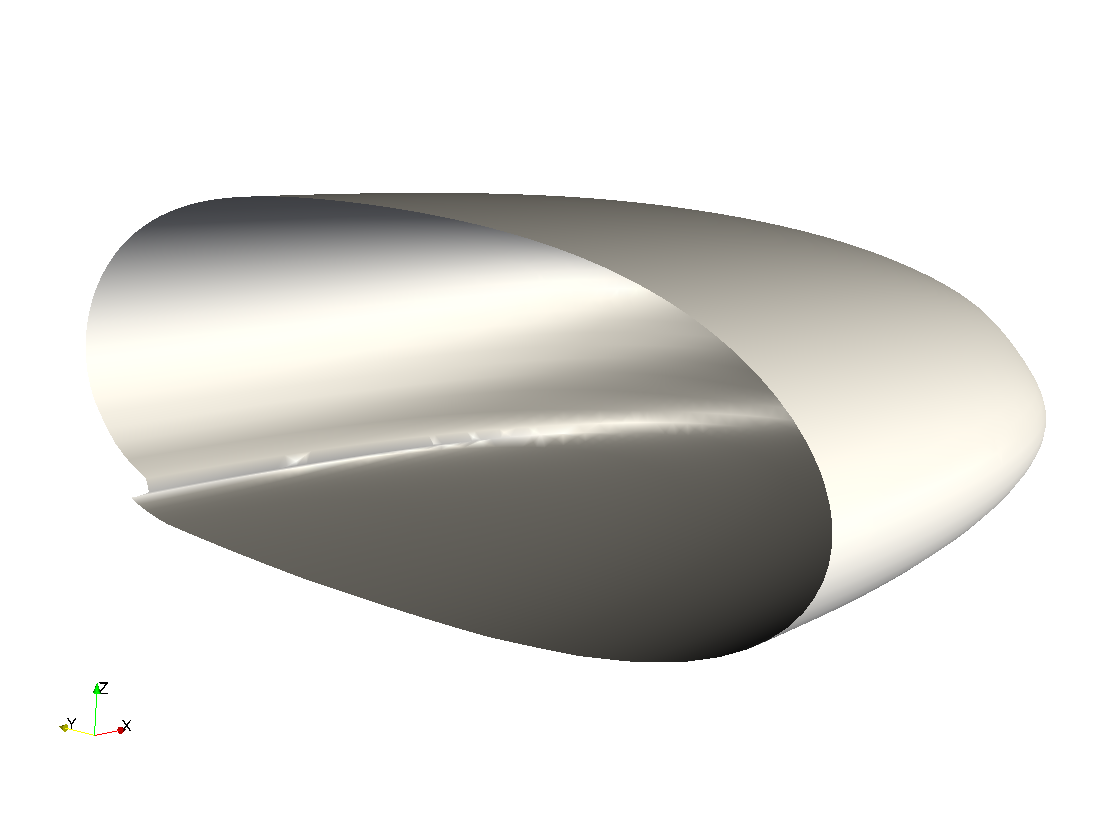}\\[2em]
  \includegraphics[width=0.49\textwidth]{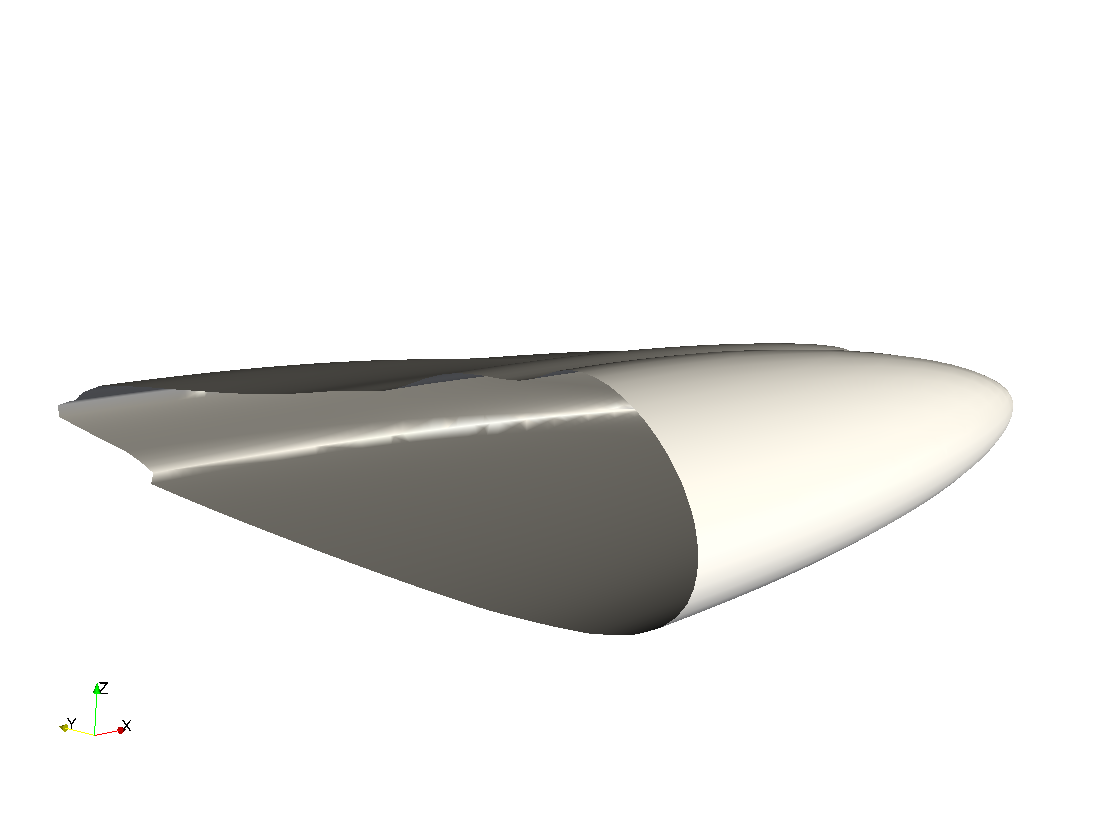}
  \includegraphics[width=0.49\textwidth]{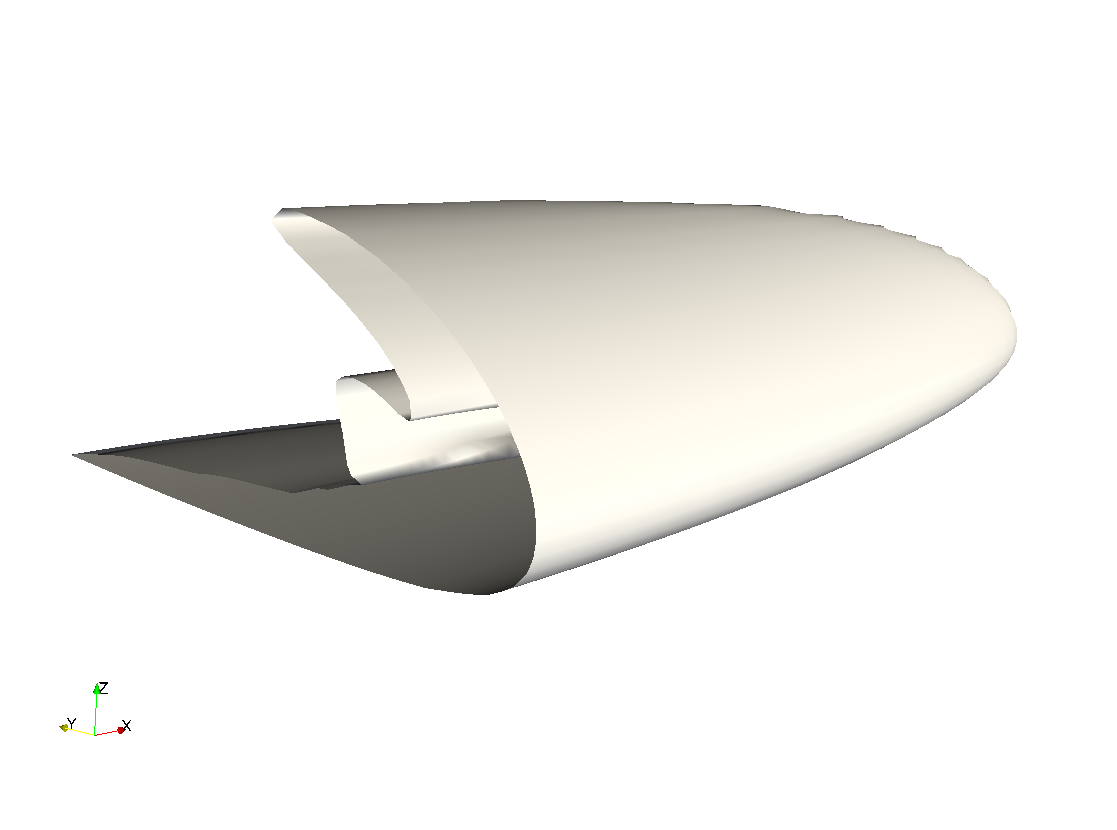}\\[4em]
  \includegraphics[width=0.49\textwidth]{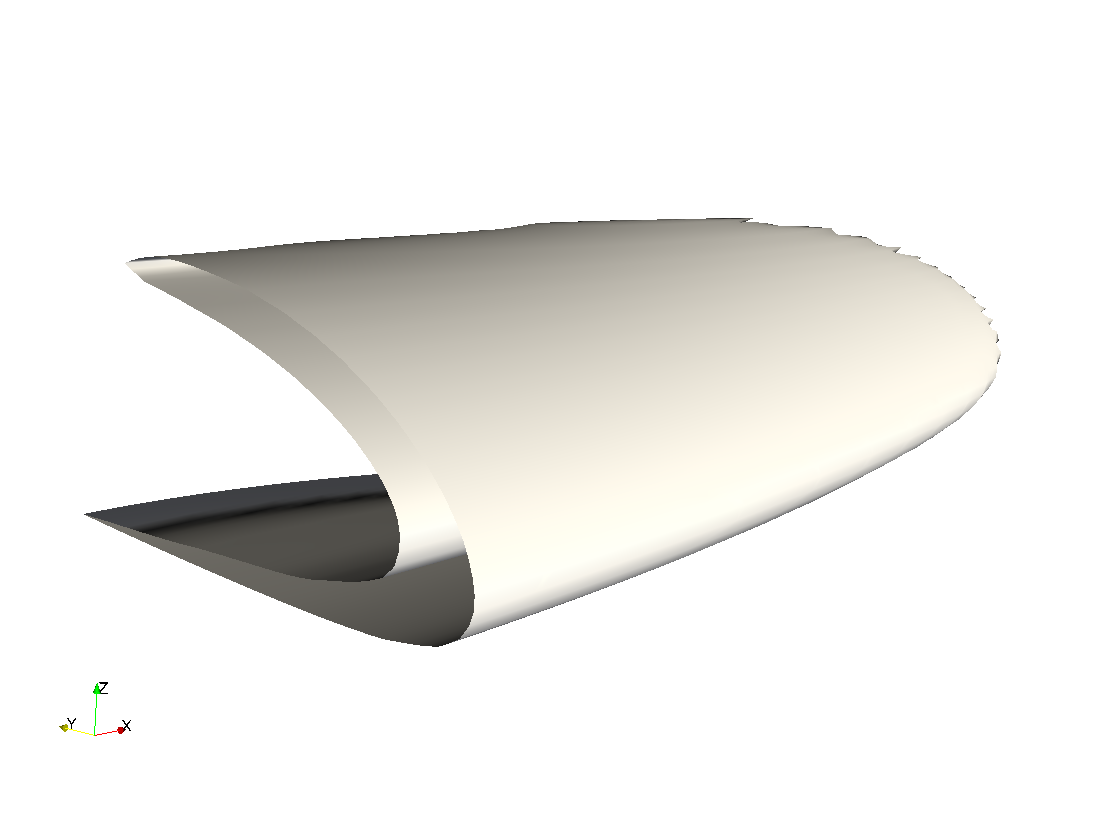}
  \includegraphics[width=0.49\textwidth]{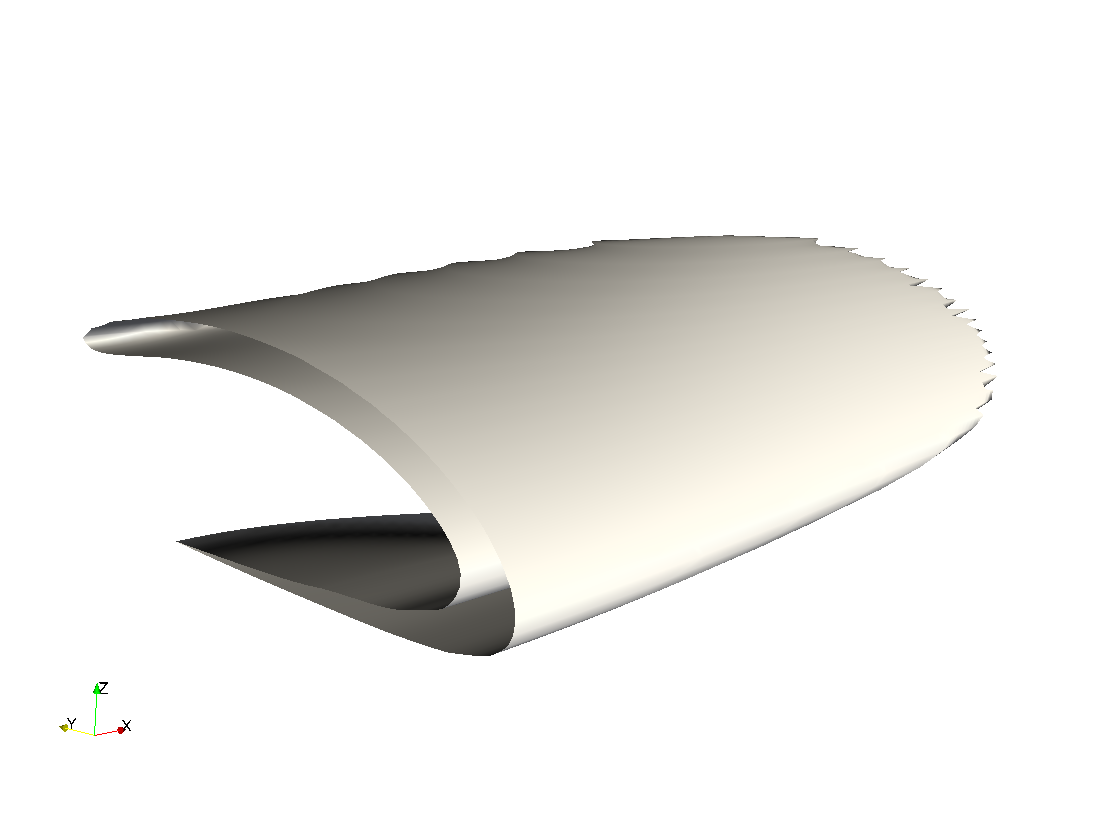}\\[2em]
  \caption[Evolution of the Lamb dodecapole]{Evolution of the Lamb
    dodecapole. Pictured are isosurfaces of the absolute vorticity,
    $|\vor(x,t)|$ at 75\% of the peak vorticity. Again, only one of
    twelve tubes is shown. As before, the vortex tube is flattens,
    followed by a roll-up. A secondary vortex sheet develops and gets
    drawn out of the center region. All pictures are from run
    \texttt{lamb}.}
  \label{fig:lambevolution}
\end{figure}

\subsubsection{Lamb dodecapole}

The Lamb dodecapole initial conditions are motivated by the fact that
each Lamb dipole in itself is an exact and invariant solution to the
Euler equations. It was therefore anticipated in
\cite{orlandi-carnevale:2007} that a more complex setup consisting of
Lamb dipoles will exhibit considerably less core deformation for the
vortex tubes. If this assumption would be met, the dodecapole
arrangement could lead to the formation of a locally self-similar
blowup scenario: The Lamb-dipoles would approach and amplify each
other, but, without core deformation, stay in their relative alignment
and shape. The ever-decreasing length-scale would result in a
point-wise collapse to the origin.

As shown in Fig. \ref{fig:lambevolution}, this scenario is not
observed in the numerical simulation. The initial tubes are deformed
severely in the course of the simulation. Vortex core deformation is
not prevented. This is hardly surprising, since the vortex dodecapole
relies on strain imposed by the rotational images of the tube
\textit{by design}, while the Lamb dipole configuration only prevents
deformation by the reflectional image. Due to the initially close
proximity of all twelve vortex tubes and the short timescale of the
evolution, deformation induced by the reflectional partner seems to be
negligible, regardless of the actual vorticity profile of the tubes.

Altogether, the evolution of the vortex tubes for the Lamb case
resembles the above presented vortex dodecapole flow: An initial
flattening of the tubes is followed by a roll-up. The emerging
secondary vortex sheet gets drawn out and finally exceeds the original
sheet in length. Due to the overall similarity of both flows it seems
safe to deduce that the topological flow evolution only weakly depends
on the precise vorticity profile. This may be seen as motivation to
transfer the results for just one particular initial condition to the
whole class of vortex dodecapole flows.
\begin{figure}[t]
  \centering
  \includegraphics[width=0.98\textwidth]{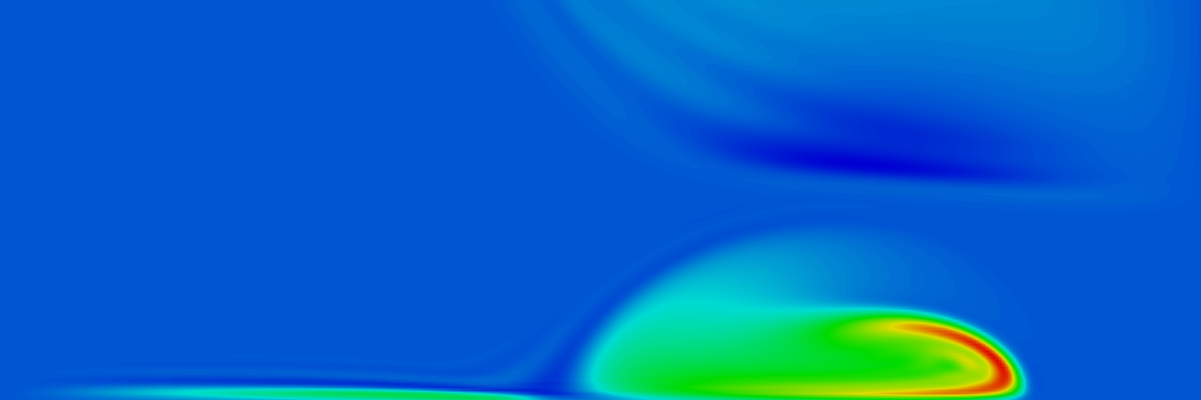}\\[1em]
  \includegraphics[width=0.98\textwidth]{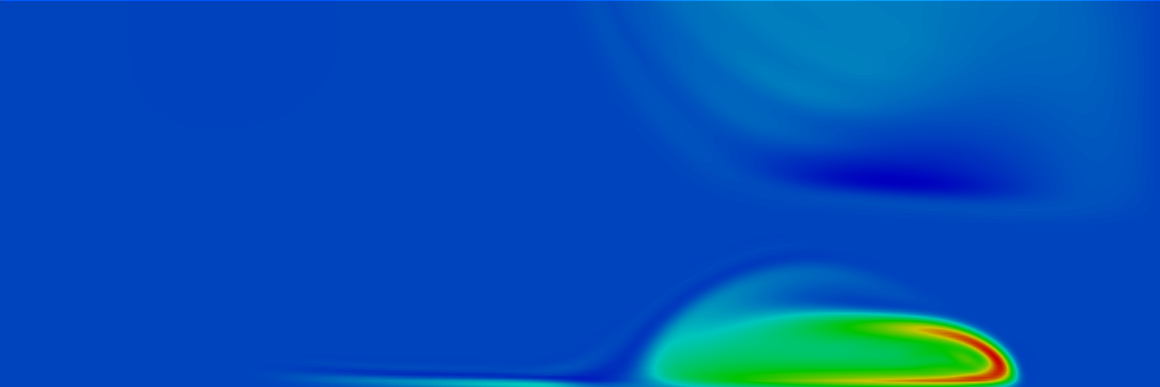}
  \caption{Direct comparison between the Lamb and the simple vorticity
    profile at late time. The vorticity in a slice near the plane of
    symmetry, $z=0.1$, is pictured. The Lamb dodecapole (top)
    exhibits a more pronounced trailing vortex sheet near the symmetry
    plane. This effect is considerably smaller for the simple
    vorticity profile (bottom).}
  \label{fig:evolutioncomparison}
\end{figure}

\subsubsection{Comparison and conclusion}

The results of the previous two sections lead to the conclusion that
no remarkable differences exist in the overall properties of the
flow. In Fig. \ref{fig:evolutioncomparison}, a direct comparison
between low resolution runs ($1024^3$) for the simple and the Lamb
vorticity profile are shown for a late time to reveal the details of
the differences. Most of the large-scale structures are identical for
both flows. The initial shape of the Lamb profile is responsible for
the formation of a less sharp roll-up of the vortex sheet and the
accumulation of secondary vorticity inside the kink. Furthermore, the
trailing vortex sheet, which is an artifact of the collapse of the
vortex dipoles to the center, is considerably stronger for the Lamb
dipoles.

Since, additionally, the core deformation is not effectively prevented
in the Lamb case, these arguments were the reason that all high
resolution runs and all geometric diagnostics were performed for the
simple, smooth dodecapole initial conditions.

\subsection{Accumulation of vorticity and strain}

The vortex dodecapole is designed to be a violent initial condition
with rapid accumulation of vorticity. Unlike e.g. the Taylor-Green
vortex or Kerr's initial conditions, no sustained phase of flow
evolution has to be awaited for the critical structures to form. Thus,
vorticity accumulation sets in immediately.

The overall vorticity amplification from initially $\Omega(0)=20$ to
finally $\Omega(t)>10^5$ exceeds a factor of $500$. The location of
the maximum vorticity follows the tip of the vortex sheet show above,
and is located at the intersection of the vortex sheets when the
roll-up begins to form.  The growth of the maximum of the norm of the
strain, $\|S(\cdot,t)\|_{L^\infty}$ behaves in a similar manner as the
peak vorticity, which is about $\|S(\cdot,0)\|_{L^\infty} \approx
12.4$ initially and grows by more than two orders of magnitude in the
course of the simulation.

\begin{figure}[t]
  \centering
  \includegraphics[width=0.9\textwidth]{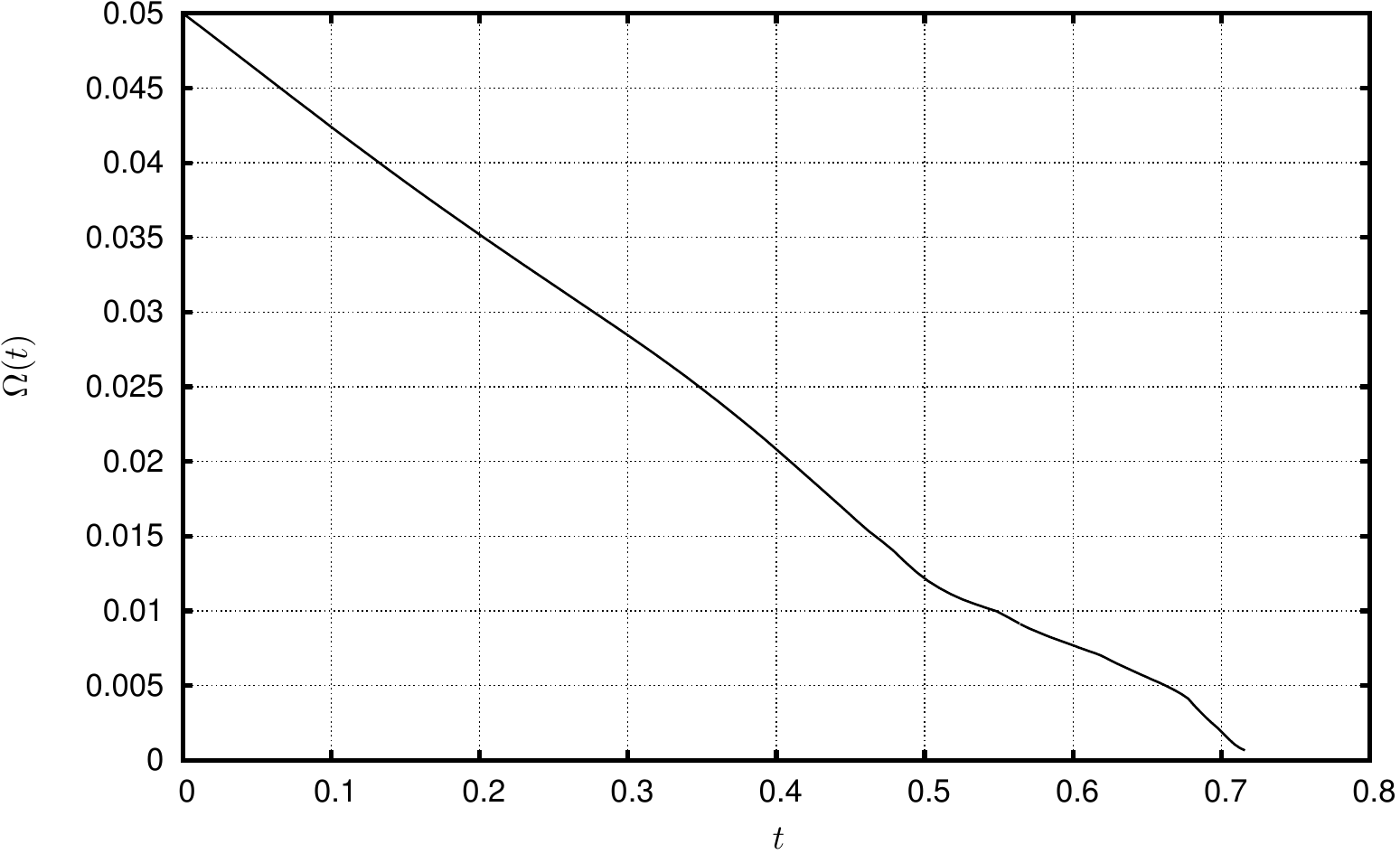}
  \caption{Evolution of $1/\Omega(t)$ in time. This mode of plotting
    suggests a growth of $\Omega(t) \approx 1/(T-t)^\gamma$ with
    $\gamma = 1$ and a blowup time $T\approx 0.72$.}
  \label{fig:bkminverse}
\end{figure}
The BKM-criterion implies that the growth in time of $\Omega(t)$ has
to fulfill $\Omega(t) \approx 1/(T-t)^\gamma$ with $\gamma \ge 1$ to
be compatible with a finite-time singularity. A plot of $1/\Omega(t)$
(by assuming $\gamma = 1$) is pictured in Fig.
\ref{fig:bkminverse}. At small times $t$, this graph looks straight,
but the growth rate changes at least twice in the evolution of the
flow. This can be explained by competing maxima in $|\vor|$ overtaking
the original $\Omega(t)$, thus changing the growth rate at different
stages. Nevertheless, at no time the vorticity looks as though
saturating, and in the latest stage of development suggests a blowup
time of $T\approx 0.72$.

Numerical data of this kind has been interpreted as evidence in favor
of the formation of a finite-time singularity before. Yet, even though
the plot \ref{fig:bkminverse} is rather suggestive, the growth may as
well be fitted to some fast (double) exponential growth.

\subsection{Geometry of the critical vortex line}

It was stated by theorem 1 of \cite{deng-hou-yu:2005} that a blowup
of vorticity in any point $\x$ is impossible as long as for some $\y$
on the same vortex line uncritical growth of vorticity is observed and
along the vortex line connecting $\x$ to $\y$ the integral of $\nabla
\cdot \vxi$ remains bounded. As lined out above, we re-interpreted
this statement as: Supposing there is singular behavior of the
maximum vorticity $\Omega(t)$, does the flow allow for a point-wise
blowup or is there a blowup of a finite, non-vanishing vortex line
segment?

Numerically, this test was implemented as follows:
\begin{itemize}
\item At each timestep, identify the point of maximum vorticity as
  $\x(t)$.
\item Follow the vorticity direction vector field while integrating
  $\nabla \cdot \vxi$ along the path. This is done with a third-order
  Runge-Kutta integrator in space.
\item As soon as the integrated quantity exceeds the threshold $C$,
  identify the current location on the vortex line as $\y(t)$. To
  increase precision, the endpoint is found via bisection.
\item Geometric properties and diagnostics for the vortex line segment
  are calculated, especially its length and $|\vor(\y(t),t)|$ to
  distinguish the cases introduced above.
\end{itemize}
This procedure is carried out for the whole time interval, as long as
the simulation is well resolved. The constant $C$ is chosen in a
reasonable way to achieve a length of the vortex line segment that
fits into the computational domain in the beginning of the simulation,
but is still well resolved at the chosen resolution at later
times. Hence, the whole vortex line segment is resolved reliably
throughout the simulation.

\begin{figure}[t]
  \centering
  \includegraphics[width=0.45\textwidth]{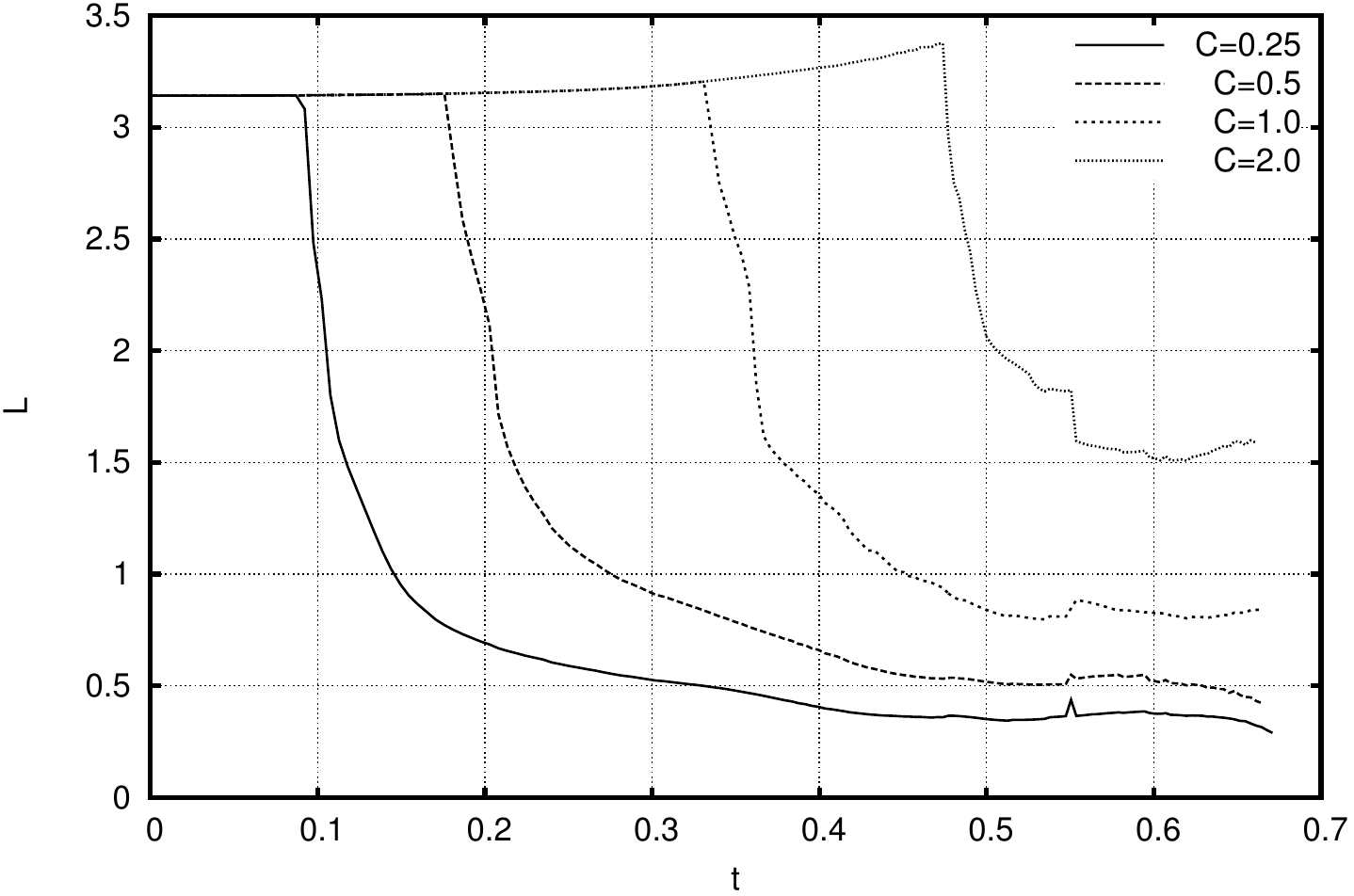}\hspace{0.05\textwidth}
  \includegraphics[width=0.45\textwidth]{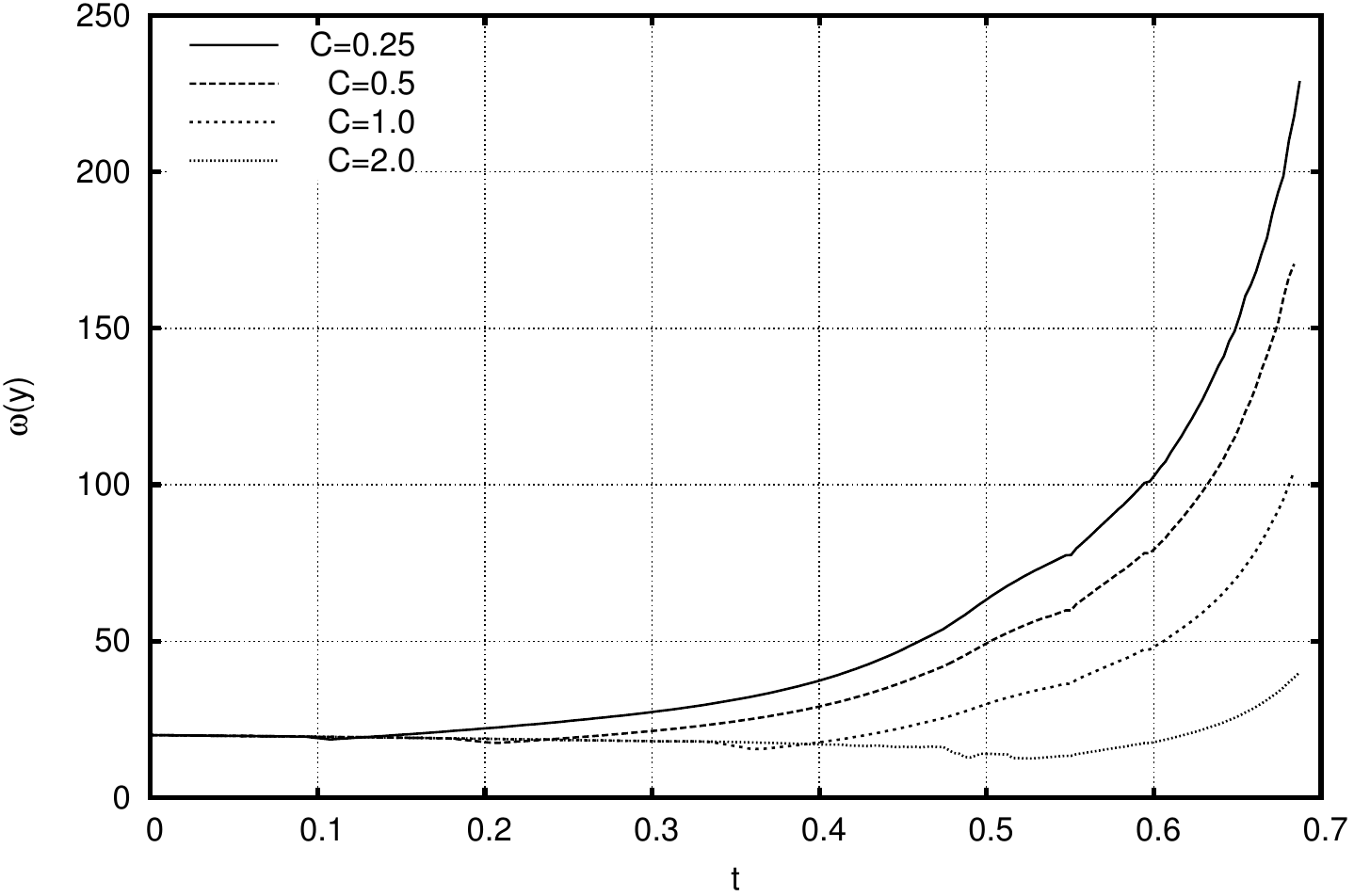}
  \caption{Left: Length of the vortex lines starting at position $\x$
    of maximum vorticity for constant $C=\int_x^y \nabla \cdot \xi
    \mathrm{d}s$. Right: Vorticity at the endpoint $y$ of these vortex
    line. Once satiated, the growth rate is the same for all $y$.}
  \label{fig:T1results}
\end{figure}

The results for the vortex dodecapole initial conditions are presented
in Fig. \ref{fig:T1results} for different constants $C \in \{0.25,
0.5, 1, 2\}$. Initially, the vortex line segments do not accumulate
enough $\nabla \cdot \vxi$, so that the length is bounded by the size
of the computational domain ($\x \in [0,\pi]^3$). At some point,
depending on the value of $C$, the threshold is reached and the length
of the vortex line segment decreases. Yet, for all considered cases of
$C$, the length does not collapse to a point, but saturates at early
times without approaching $l(t) = 0$. This behavior appears to be
stable up to the latest time of the simulation. The final length of
the vortex line segments is at least $0.3$ for the smallest case of
$C$ ($C=0.25$), which is still well resolved with at least
$200\,\Delta x$. This result, therefore, is a numerical evidence
against a point-wise blowup for the vortex dodecapole class of initial
conditions. This is in concordance with the estimate in
\cite{deng-hou-yu:2005}.

Monitoring the development of $\vor(\y(t),t)$ yields, as shown in
Fig. \ref{fig:T1results} (right), a similar growth rate for the
accumulation of vorticity at the endpoint as for the beginning of the
vortex line segment. This is hardly surprising, since by construction
a constant value for $C$ directly links the growth rates of
$|\vor(\x(t),t)|$ to $|\vor(\y(t),t)|$. Nevertheless, a numerical
verification of this analytic equality may be seen as a confirmation
that the observed growth rate of $|\vor(\x(t),t)|$ is by no means a
numerical artifact in an isolated small area, but is reproduced at
points far away from the critical region, which appear to be
well-behaved at first view. The possibly critical growth in the
perspective of BKM is, thus, confirmed by the global flow.

Furthermore, since for a large portion of the simulation the distance
$l(t)$ is approximately constant, this could possibly be seen as an
evidence for the existence of a non-vanishing vortex line segment that
blows up in every point. The popular scenario of a collapse to a
single point, on the other hand, is clearly conflicting the numerical
evidence. The discovery of a possibly critical vortex line segment in
the vortex dodecapole flow, however, is to be handled with care,
since distinguishing between critical and sub-critical blowup of the
whole segment is in no way more conclusive than distinguishing between
critical and sub-critical growth of $\Omega(t)$. Thus, learning from
the lesson taught by 25 years of numerically testing BKM, this should
not be interpreted as clear evidence in favor of a finite-time
singularity.

\subsection{Lagrangian evolution of the critical vortex line segments}

The geometric properties of Lagrangian vortex line segments,
especially their curvature $\kappa$ and the tightening of their
surroundings $\nabla \cdot \vxi$ have been established as revealing
parameters in understanding the nature of rapid accumulation of
vorticity in Euler flows and a sound connection to singular behavior
is made through theorem 2. The ambition here is to utilize these
geometric properties, monitored in a numerical simulation, as more
reliable means of distinguishing between a finite-time singularity and
a mere fast accumulation of vorticity.

Despite high hopes from an analytical point of view that these
considerations will shed light on the nature of vorticity
accumulation, numerical results observing geometrical properties of
Lagrangian vortex filaments are scarce. This is primarily due to the
fact that Eulerian quantities such as $\Omega(t)$ are readily
trackable in post-processing, while monitoring the Lagrangian
evolution requires additional computational effort. On top of that,
the geometry of integral curves at an instance in time, though in
principle computable in post-processing, as well as derived quantities
such as their convergence and curvature, are quite inaccessible in
comparison to simple Eulerian criteria.

This section is devoted to the presentation of results concerning the
assumptions of theorem 2 of \cite{deng-hou-yu:2005} for the vortex
dodecapole initial conditions. Quite similar to the first theorem,
there is considerable freedom in the choice of the involved
quantities. The strategy we chose in the context of this paper is as
follows:
\begin{itemize}
\item Identify the Lagrangian fluid element $\Alpha$, which will
  contain the maximum of vorticity at the latest time of the
  simulation, $\Omega(t) \approx |\vor(\X(\Alpha,t),t)|$. A vortex
  line segment $L_t$ starting here will intrinsically be
  ``comparable'' to the maximum of vorticity (as in
  $|\vor(\X(\Alpha,t),t)| \gtrsim \Omega(t)$) at late stages of the
  simulation. The assumptions concerning the segment are therefore
  automatically met. In the numerics this procedure is implemented by
  carrying out a precursory identical simulation with a huge number of
  tracer particles ($\approx$ 1 million) randomly distributed across
  the domain. Particles that accumulate huge amounts of vorticity are
  selected for the subsequent production run.
\item For the production run, at each instance in time start a vortex
  line integration at $\X(\Alpha,t)$ along the vorticity direction
  field. Monitor the maximum curvature $\|\kappa\|_{L^\infty(L_t)}$
  and the maximum vortex line convergence $\|\nabla \cdot
  \vxi\|_{L^\infty(L_t)}$ during the integration and calculate
  $\lambda(t)$. Stop the integration, as soon as $\lambda(t)$ reaches
  a fixed, arbitrary constant $C$. This defines $L_t$. In the numerics
  this is again implemented with a third-order Runge-Kutta integration
  and bisectioning to obtain the endpoint of $L_t$.
\item For this vortex line segment $L_t$, calculate the length $l(t)$,
  and the velocity components $U_n$ and $U_\xi$. From the collapse of
  the length $l(t)$ approximate the exponent $B$. This in turn
  provides the critical growth exponent $A$ for the velocity
  variables, $A_{\text{crit}} = 1 - B$.
\item Compare the increase in $U_n$ and $U_\xi$ to
  $1/(T-t)^{A_{\text{crit}}}$ to distinguish between critical and
  sub-critical growth of velocity.
\end{itemize}

This can be interpreted rather intuitively. By prescribing an
arbitrarily fixed $\lambda(t)$, the vortex line segment is kept
relatively geometrically uncritical, as the length-scale is always
adjusted accordingly. This process of ``zooming in'' just enough to
retain the geometric ``criticalness'' prescribes the rate of collapse
to a point, at least in the direction of the vortex line. All that is
left to check is whether the velocity growth in the immediate
surrounding is fast enough to be compatible with a finite-time
singularity.

The results of the previous section, concerning a point-wise
singularity versus the blowup of a whole vortex segment, already
anticipates that the increase in $\nabla \cdot \vxi$ around the
critical vortex line is bounded. If the curvature of the vortex line
segment remains controllable (which is to be expected from the
pictures), then just a mild collapse of $l(t)$ occurs. This leaves
much room for $U_n$ and $U_\xi$ to still be distinguishable from a
critical growth.

\begin{figure}[t]
  \centering
  \includegraphics[width=0.45\textwidth]{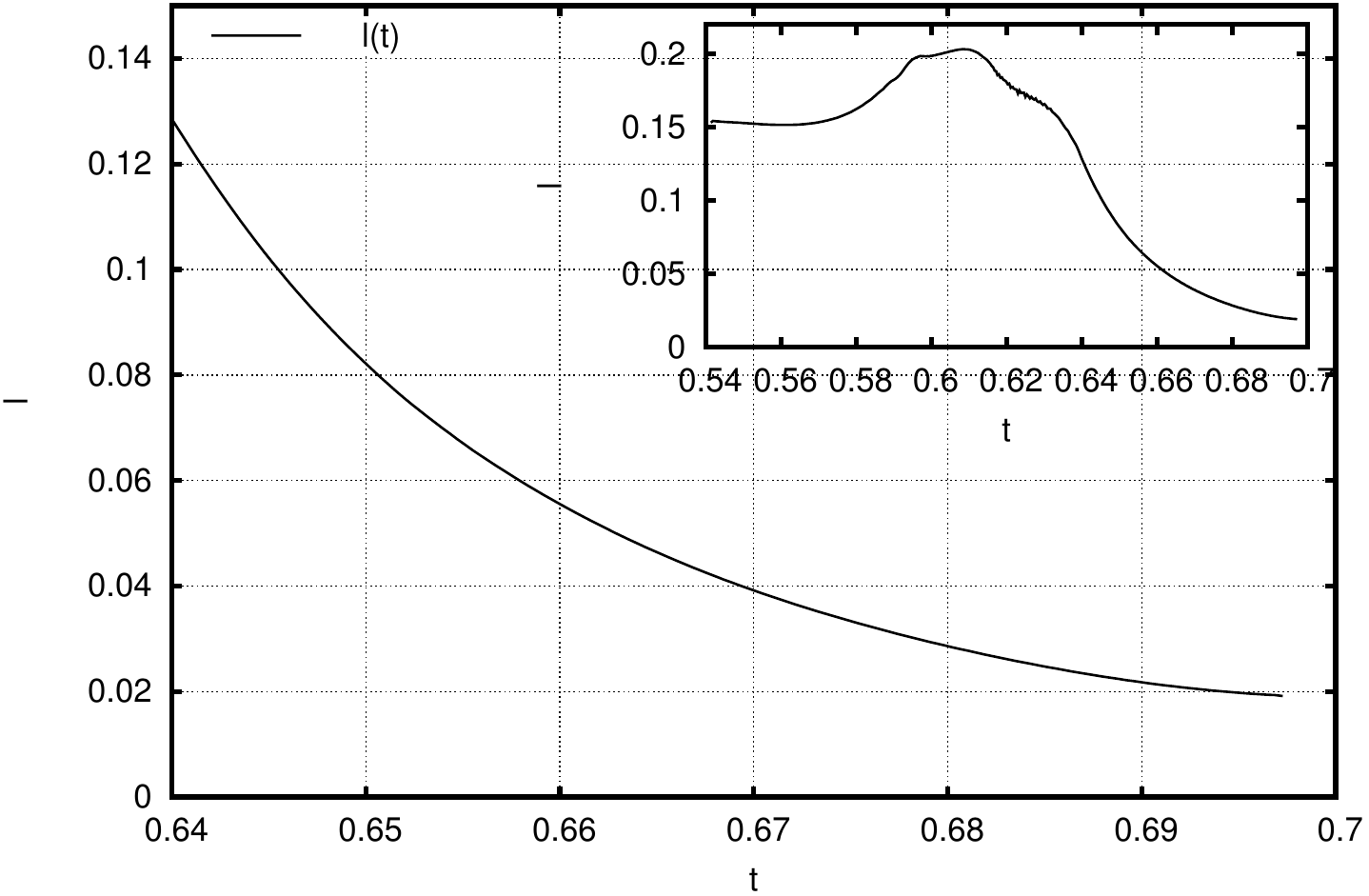}\hspace{0.05\textwidth}
  \includegraphics[width=0.45\textwidth]{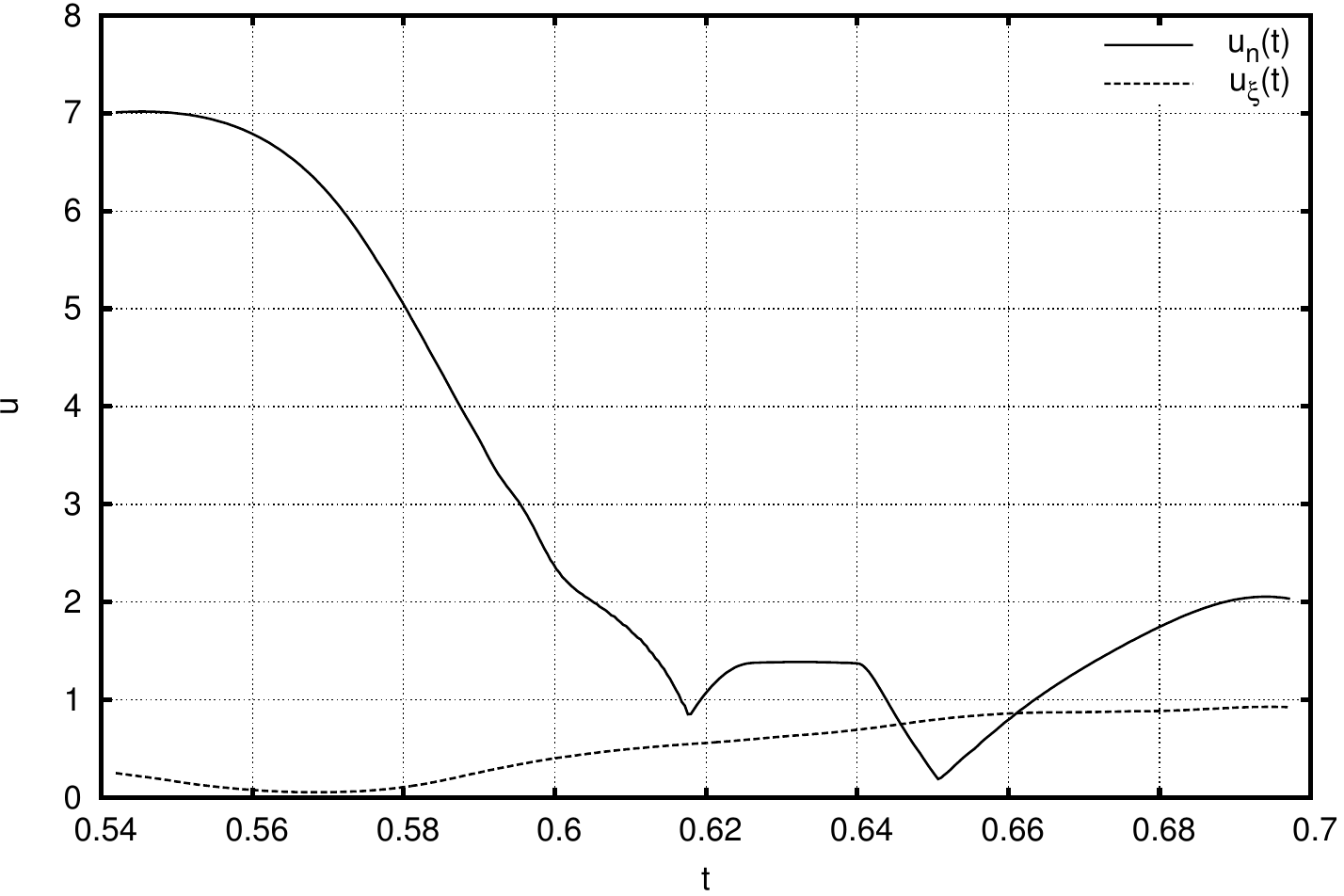}
  \caption{Top: Evolution of the length $l(t)$ of the critical vortex
    filament $L_t$ for different Lagrangian fluid elements. The length
    does not decrease as $(T-t)^B$ for any $B < 1$, which would be
    faster than linear. The Lagrangian collapse of the vortex segment
    is decelerating instead. Bottom: Evolution of the quantities $U_n$
    and $U_\xi$ in time. $U_n$ does not appear to be growing, while
    $U_\xi$, though increasing in time, does not exhibit a finite-time
    blowup as $1/(T-t)$.}
  \label{fig:T2results}
\end{figure}
Fig. \ref{fig:T2results} shows the results for the vortex dodecapole
initial conditions for a fixed constant $C$. Different choices of the
constant $C$ produce identical, rescaled results. The top plot
pictures the length of the vortex line segment for the tracer that is
arriving at a position of very huge vorticity at late stages of the
simulation. The subplot depicts the long-term behavior of the particle
entering the critical region, while the final stage of length decrease
is magnified. The decrease in length does not agree with a collapse in
final time, but instead the shrinkage of the segment decelerates
clearly in time. This contradicts a scaling in time proportional to
$(T-t)^B$ for any $0 < B \le 1$, which would be faster than (or, in
the limiting case, equal to) linear. It should be noted that for the
observed collapse in length, the vortex segment curvature $\kappa$ is
the dominating term in $M(t) = \max(\|\nabla \cdot
\vxi\|_{L^\infty(L_t)},\|\kappa\|_{L^\infty(L_t)})$, shadowing the
effects of $\nabla \cdot \vxi$. This may lead to a change of regime in
the rate of collapse, if $\nabla \cdot \vxi$ at some point exceeds
$\kappa$ in quantity.

It could furthermore be argued that the limit $B \rightarrow 0$ is
hard to exclude, since the drop in length would be virtually
instantaneous in time, with a close to constant scaling before. In
this limit, the quantities $U_n$ and $U_\xi$ would have to grow
roughly as $1/(T-t)$ to still allow formation of a finite-time
singularity. $U_\xi$ quantifies the largest difference in axial
velocity along the segment. For an isolated collapsing vortex tube,
this quantity can be expected to not increase critically, since the
tangential velocity is less likely to rapidly change than the radial
velocity. However, this initial explanation does not consider the
influence of the mirror tubes. $U_n$ on the other hand may be
interpreted as the velocity of the vortex tube's core itself. Again,
in an isolated setup this velocity is not expected to blow up.

Fig. \ref{fig:T2results} (bottom) shows the observed behavior of $U_n$
and $U_\xi$ in time for the tracer as considered above. $U_n$ stays
roughly constant in time, showing no signs of a blowup. $U_\xi$, even
though increasing in time, does not fit to critical growth, in
particular not like $1/(T-t)$ in time. Thus, the assumptions of
theorem 2 are well met. This result therefore poses a strong numerical
evidence against a finite-time singularity for the class of vortex
dodecapole initial conditions.

\section{Conclusion}

In this paper we present numerical evidence against the formation of a
finite-time singularity for the vortex dodecapole initial
condition. We use data obtained from high resolution adaptively
refined numerical simulations to test the assumptions presented by
geometric blowup criteria. The applied numerical method allows for a
clearer insight into the formation of the possible singularity. Most
notably, it implies numerical techniques to distinguish between a
point-wise blowup and the blowup of a whole vortex line
segment. Furthermore, by tracking curvature and spreading of
Lagrangian vortex line segments, the distinction between singular and
non-singular behavior can be made much more clearly than the usual
approach via BKM .

In this paper we used vortex dodecapole initial conditions with two
different vorticity profiles. Comparison of the simulation shows that
different vorticity profiles yield similar visual and geometrical
appearance. This serves as an argument that the obtained results may
apply to the whole class of vortex dodecapole flows.  Monitoring the
growth rate of $\Omega(t)$ quantifies the well-known vorticity
amplification. Amplification by more than two orders of magnitude was
reached for both vorticity and strain, exceeding by far values
achieved by previous simulations
\citep{grafke-homann-dreher-grauer:2008}. Applying this data to BKM
would lead to the conclusion that a finite-time singularity at time
$T\approx0.72$ fits via extrapolation. Yet, as the history of Euler
simulations has shown, statements obtained by extrapolation are to be
handled with care.

Following the presented argument, a point-wise collapse should
coincide with a blowup of $\nabla \cdot \vxi$ at the point of maximum
vorticity. As shown, this statement can be broadened: A finite-time
singularity must either lead to a blowup of $\nabla \cdot \vxi$ at the
point of maximum vorticity, or the whole critical vortex line segment
has to blow up. Utilizing the geometric information obtained via
vortex line integration from the numerical simulation, it is observed
that $\nabla \cdot \vxi$ does not grow in a way to be compatible with
a point-wise collapse. Yet, measuring the growth-rates on the critical
vortex line, high rates of amplification are measured far away from
the critical region. Even though it is hard to distinguish, whether
this amplification is critical or sub-critical, this might be
interpreted as an evidence for the blowup of the complete vortex line
segment, even though it suffers exactly the same vulnerabilities as
extrapolation in BKM. A point-wise blowup, on the other hand, cleary
contradicts the numerical results up to the time reached.

Evidence against a blowup of the whole critical vortex line segment is
found when looking at the geometric properties of Lagrangian vortex
line segments. It was shown by theorem 2 in \citep{deng-hou-yu:2005}
that a blowup of vorticity is directly connected to the interplay
between velocity growth and the collapse of vortex line filaments,
when maintaining the overall same shape in geometric means (i.e. the
same $\lambda(L_t)$). Since curvature and $\nabla \cdot \vxi$ do not
increase in order to support a finite-time collapse of the segment,
velocity components in the vicinity of the vortex line filament would
have to increase as $1/(T-t)$ to support the blowup hypothesis. Up to
the time reached, this critical velocity growth may be excluded by
numerical means. This poses a numerical evidence against the formation
of a singularity in finite time for vortex dodecapole configurations.

\section*{Acknowledgment}

We would like to thank J. Dreher for his work on the computational
framework.  This work benefited from support through project
\mbox{GR~967/3-1} of the Deutsche Forschungsgesellschaft. Access to
the BlueGene/P multiprocessor computer JUGENE at the Forschungszentrum
J\"ulich was made available through project hbo35.


\begin{thebibliography}{10}
\expandafter\ifx\csname url\endcsname\relax
  \def\url#1{\texttt{#1}}\fi
\expandafter\ifx\csname urlprefix\endcsname\relax\def\urlprefix{URL }\fi
\expandafter\ifx\csname href\endcsname\relax
  \def\href#1#2{#2} \def\path#1{#1}\fi

\bibitem{fefferman:2000}
C.~Fefferman, Existence and smoothness of the {N}avier-{S}tokes equation,
  published online:\\
  {\footnotesize\texttt{http://www.claymath.org/millennium/}} (2000).

\bibitem{ladyzhenskaya:2003}
O.~A. Ladyzhenskaya, Sixth problem of the millenium: {N}avier-{S}tokes
  equations, existence and smoothness, Russian Math. Surveys 58 (2001)
  251--286.

\bibitem{doering:2009}
C.~R. Doering, The 3{D} {N}avier-{S}tokes problem, Annual Review of Fluid
  Mechanics 41 (2009) 109--128.

\bibitem{caffarelli-kohn-nirenberg:1982}
L.~Caffarelli, R.~Kohn, L.~Nirenberg, Partial regularity of suitable weak
  solutions of the {N}avier-{S}tokes equations, Comm. Appl. Math. 35 (1982)
  771--831.

\bibitem{leray:1934}
J.~Leray, Sur le mouvement d'un liquide visqeux emplissant l'espace, Acta Math.
  63 (1934) 193--248.

\bibitem{constantin:2007}
P.~Constantin, On the {E}uler equations of incompressible fluids, Bulletin
  Amer. Math. Soc. 44~(4) (2007) 603--621.

\bibitem{she-leveque:1994}
Z.~She, E.~L\'ev\^eque, Universal scaling laws in fully developed turbulence,
  Physical Review Letters 72~(3) (1994) 336--339.

\bibitem{frisch:1995}
U.~Frisch, Turbulence, Cambridge University Press, Cambridge, 1995.

\bibitem{kolmogorov:1941}
A.~N. Kolmogorov, Local structure of turbulence in an incompressible fluid at
  very high {R}eynolds numbers, Dokl. Akad. Nauk SSSR 30 (1941) 299--303.

\bibitem{kolmogorov:1941b}
A.~N. Kolmogorov, Energy dissipation in locally isotropic turbulence, Dokl.
  Akad. Nauk SSSR 32 (1941) 19--21.

\bibitem{kolmogorov:1962}
A.~N. Kolmogorov, A refinement of previous hypotheses concerning the local
  structure of turbulence in a viscous incompressible fluid at high {R}eynolds
  numbers, J. Fluid Mech. 13 (1962) 82--85.

\bibitem{onsager:1949}
L.~Onsager, Statistical hydrodynamics, Nuovo Cimento (Supplemento) 6 (1949)
  279--287.

\bibitem{constantin-e-titi:1994}
P.~Constantin, W.~E, E.~S. Titi, Onsager's conjecture on the energy
  conservation for solutions of {E}uler's equation, Commun. Math. Phys. 165
  (1994) 207--209.

\bibitem{cheskidov-constantin-friedlander-shvydkoy:2008}
A.~Cheskidov, P.~Constantin, S.~Friedlander, R.~Shvydkoy, Energy conservation
  and {O}nsager's conjecture for the {E}uler equations, Nonlinearity 21~(6)
  (2008) 1233.

\bibitem{gibbon:2008}
J.~D. Gibbon, The three-dimensional {E}uler equations: Where do we stand?,
  Physica D: Nonlinear Phenomena 237 (2008) 1894--1904.

\bibitem{beale-kato-majda:1984}
J.~T. Beale, T.~Kato, A.~Majda, Remarks on the breakdown of smooth solutions
  for the 3-{D} {E}uler equations, Commun. Math. Phys. 94 (1984) 61--66.

\bibitem{constantin-fefferman-majda:1996}
P.~Constantin, C.~Fefferman, A.~Majda, Geometric constraints on potentially
  singular solutions for the 3{D} {E}uler equations, Commun. Part. Diff. Eq. 21
  (1996) 559--571.

\bibitem{cordoba-fefferman:2001}
D.~Cordoba, C.~Fefferman, On the collapse of tubes carried by 3{D}
  incompressible flows, Commun. Math. Phys. 222 (2001) 293--298.

\bibitem{deng-hou-yu:2005}
J.~Deng, T.~Y. Hou, X.~Yu, Geometric properties and nonblowup of 3{D}
  incompressible {E}uler flow, Commun. Part. Diff. Eq. 30~(1-2) (2005)
  225--243.

\bibitem{constantin:1994}
P.~Constantin, Geometric statistics in turbulence, SIAM Rev. 36.

\bibitem{gibbon:2002}
J.~D. Gibbon, A quaternionic structure in the three-dimensional {E}uler and
  ideal magneto-hydrodynamics equation, Physica D: Nonlinear Phenomena
  166~(17).

\bibitem{gibbon-holm-kerr-roulstone:2006}
J.~D. Gibbon, D.~D. Holm, R.~M. Kerr, I.~Roulstone, Quaternions and particle
  dynamics in {E}uler fluid flow, Nonlinearity 19~(1969).

\bibitem{deng-hou-yu:2006}
J.~Deng, T.~Y. Hou, X.~Yu, Improved geometric conditions for non-blowup of the
  3{D} incompressible {E}uler equation, Commun. Part. Diff. Eq. 31~(2) (2006)
  293--306.

\bibitem{ashurst-kerstein-kerr-gibson:1987}
W.~T. Ashurst, A.~R. Kerstein, R.~M. Kerr, C.~H. Gibson, Alignment of vorticity
  and scalar gradient with strain rate in simulated {N}avier–{S}tokes
  turbulence, Phys. Fluids 30 (1987) 2343.

\bibitem{meneveau:2011}
C.~Meneveau, Lagrangian dynamics and models of the velocity gradient tensor in
  turbulent flows, Annual Review of Fluid Mechanics 43 (2011) 219--245.

\bibitem{chevillard-meneveau:2011}
L.~Chevillard, C.~Meneveau, Lagrangian time correlations of vorticity
  alignments in isotropic turbulence: Observations and model predictions, Phys.
  Fluids 23.

\bibitem{pumir-siggia:1990}
A.~Pumir, E.~Siggia, Collapsing solutions to the 3-{D} {E}uler equations, Phys.
  Fluids A 2~(2) (1990) 220--241.

\bibitem{hamlington-schumacher-dahm:2008}
P.~E. Hamlington, J.~Schumacher, W.~J.~A. Dahm, Direct assessment of vorticity
  alignment with local and nonlocal strain rates in turbulent flows, Phys.
  Fluids 20.

\bibitem{taylor-green:1937}
G.~I. Taylor, A.~E. Green, Mechanism of the production of small eddies from
  large ones, Proc. R. Soc. Lond. A 158 (1937) 499--521.

\bibitem{kerr:1993}
R.~M. Kerr, Evidence for a singularity of the three-dimensional, incompressible
  {E}uler equations, Phys. Fluids A 5~(7) (1993) 1725--1746.

\bibitem{kerr:2005}
R.~M. Kerr, Velocity and scaling of collapsing {E}uler vortices, Phys. Fluids
  17.

\bibitem{hou-li:2006}
T.~Y. Hou, R.~Li, Dynamic depletion of vortex stretching and non-blowup of the
  3-{D} incompressible {E}uler equations, Journal of Nonlinear Science 16
  (2006) 639--664.

\bibitem{pelz:2001}
R.~B. Pelz, Symmetry and the hydrodynamic blow-up problem, J. Fluid Mech. 444
  (2001) 299--320.

\bibitem{kida:1985}
S.~Kida, Three-dimensional periodic flows with high symmetry, J. Phys. Soc.
  Jpn. 54 (1985) 2132--2136.

\bibitem{boratav-pelz:1994b}
O.~N. Boratav, R.~B. Pelz, Evidence for a real-time singularity in
  hydrodynamics from time series analysis, Phys. Rev. Lett. 79 (1994)
  4998--5001.

\bibitem{pelz:2003}
R.~B. Pelz, Extended series analysis of full octahedral flow: numerical
  evidence for hydrodynamic blowup, Fluid Dyn. Research 33 (2003) 207--221.

\bibitem{cichowlas-brachet:2005}
C.~Cichowlas, M.~Brachet, Evolution of complex singularities in {K}ida-{P}elz
  and {T}aylor-{G}reen inviscid flows, Fluid Dyn. Research 36 (2005) 239--248.

\bibitem{ng-bhattacharjee:1996}
C.~S. Ng, A.~Bhattacharjee, Sufficient condition for a finite-time singularity
  in a high-symmetry {E}uler flow: Analysis and statistics, Phys. Rev. E 54
  (1996) 1530--1534.

\bibitem{boratav-pelz:1994}
O.~N. Boratav, R.~B. Pelz, Direct numerical simulation of transition to
  turbulence from a high-symmetry initial condition, Phys. Fluids 6~(8) (1994)
  2757--2784.

\bibitem{grafke-homann-dreher-grauer:2008}
T.~Grafke, H.~Homann, J.~Dreher, R.~Grauer, Numercial simulations of possible
  finite time singularities in the incompressible {E}uler equations: comparison
  of numerical methods, Physica D: Nonlinear Phenomena 237 (2008) 1932--1936.

\bibitem{orlandi-carnevale:2007}
P.~Orlandi, G.~F. Carnevale, Nonlinear amplification of vorticity in inviscid
  interaction of orthogonal {L}amb dipoles, Phys. Fluids 19.

\bibitem{lamb:1932}
H.~Lamb, Hydrodynamics, Cambridge University Press, 1932.

\bibitem{wu-ma-zhou:2006}
J.-Z. Wu, H.-Y. Ma, M.-D. Zhou, Vorticity and Vortex Dynamics, Springer, 2006.

\bibitem{frisch-matsumoto-bec:2003}
U.~Frisch, T.~Matsumoto, J.~Bec, Singularities of {E}uler flow? {N}ot out of
  the blue!, J. Stat. Phys. 113 (2003) 761--781.

\bibitem{morf-orszag-frisch:1980}
R.~H. Morf, S.~A. Orszag, U.~Frisch, Spontaneous singularity in
  three-dimensional, inviscid, incompressible flow, Physical Review Letters
  44~(9) (1980) 572--575.

\bibitem{chorin:1981}
A.~J. Chorin, Estimates of intermittency, spectra, and blow-up in developed
  turbulence, Communications on Pure and Applied Mathematics 34~(6) (1981)
  853--866.

\bibitem{chorin:1982}
A.~J. Chorin, The evolution of a turbulent vortex, Commun. Math. Phys. 83
  (1982) 517--535.

\bibitem{siggia:1985}
E.~D. Siggia, Collapse and amplification of a vortex filament, Phys. Fluids 28
  (1985) 794--805.

\bibitem{bell-marcus:1992}
J.~B. Bell, D.~L. Marcus, Vorticity intensification and transition to
  turbulence in three-dimensional {E}uler equations, Commun. Math. Phys. 147
  (1992) 371--394.

\bibitem{brachet-meiron-orszag-nickel-etal:1983}
M.~E. Brachet, D.~I. Meiron, S.~A. Orszag, B.~G. Nickel, R.~H. Morf, U.~Frisch,
  Small-scale structure of the {T}aylor-{G}reen vortex, J. Fluid Mech. 130
  (1983) 411--452.

\bibitem{brachet-meneguzzi-vincent-politano-etal:1992}
M.~E. Brachet, M.~Meneguzzi, A.~Vincent, H.~Politano, P.~L. Sulem, Numerical
  evidence of smooth self-similar dynamics and possibility of subsequent
  collapse for three-dimensional ideal flows, Phys. Fluids 4~(12) (1992)
  2845--2854.

\bibitem{grauer-marliani-germaschewski:1998}
R.~Grauer, C.~Marliani, K.~Germaschewski, Adaptive mesh refinement for singular
  solutions of the incompressible {E}uler equations, Physical Review Letters
  80~(19) (1999) 4177--4180.

\bibitem{bustamante-brachet:2011}
M.~D. Bustamante, M.~Brachet, On the interplay between the {B}{K}{M} theorem
  and the analyticity-strip method to investigate numerically the
  incompressible {E}uler singularity problem, published online,
  {\footnotesize\texttt{http://arxiv.org/abs/1112.1571}} (2011).

\bibitem{dreher-grauer:2005}
J.~Dreher, R.~Grauer, Racoon: A parallel mesh-adaptive framework for hyperbolic
  conservation laws, Parallel Computing (2005) 913--932.

\bibitem{shu-osher:1988}
C.~Shu, S.~Osher, Efficient implementation of essentially non-oscillatory
  shock-capturing schemes, Journal of Computational Physics 77 (1988) 439--471.

\bibitem{kurganov-levy:2000}
A.~Kurganov, D.~Levy, A third-order semidiscrete central scheme for
  conservation laws and convection-diffusion equation., Jour. Sci. Comp. 22~(4)
  (2000) 1461--1488.

\end{thebibliography}
\end{document}